\documentstyle[epsf,epsfig,rotate]{mn}
\newcommand{\be}{\begin{equation}}
\newcommand{\ee}{\end{equation}}
\newcommand{\gs}{\;\raisebox{-.8ex}{$\buildrel{\textstyle>}\over\sim$}\;}

\newcommand{\apj}{{\it ApJ, }}

\newcommand{\icar}{{\it Icarus, }}

\title
[Protoplanets and MHD turbulence]
{The interaction of planets with a disc with MHD turbulence III: 
Flow morphology and conditions for gap formation in local and global simulations}

\author[J.C.B.Papaloizou, R.P. Nelson \& M.D. Snellgrove]{
John C.B. Papaloizou, Richard P. Nelson  \& Mark D. Snellgrove \\
Astronomy Unit, Queen Mary, University of London, Mile End Rd, London E1 4NS}

\date{Received/Accepted}

\begin{document}

\maketitle

\begin{abstract} 
We present the results of both global cylindrical disc simulations
and local shearing box simulations
of protoplanets interacting with
a disc undergoing MHD turbulence with zero net flux magnetic fields.
We investigate the nature of the disc response and conditions
for gap formation. This issue is an important one
for determining the type and nature of the migration of the protoplanet,
with the presence of a deep gap being believed to enable slower migration.

For both  types of simulation we find a common pattern of behaviour
for  which the main parameter determining the nature of the response is
$M_p R^3/(M_* H^3),$ with $M_p, M_*, R,$ and $H$ being the protoplanet mass,
the central mass, the orbital radius and the disc semi--thickness respectively.
We find that as $M_p R^3/(M_* H^3)$ is increased to $\sim 0.1$
the presence of the protoplanet is first indicated by the appearance
of the well known trailing wake
which, although it may appear to be erratic
on account of the turbulence, appears to be well defined.
Once $M_p R^3/(M_* H^3)$ exceeds  a number around
unity a gap starts to develop inside
which the magnetic energy density
tends to be concentrated in the high density wakes.
This condition for gap formation can be understood from
simple dimensional considerations of the conditions
for nonlinearity and  the balance of  angular momentum transport
due to Maxwell and Reynolds' stresses  with that due to tidal torques
applied to the parameters of our simulations.

An important result is that the basic flow morphology in the vicinity of the
protoplanet is very similar in both the local and global simulations.
This indicates that, regardless of potentially unwanted
effects arising from the periodic boundary conditions, 
local shearing box simulations,
which are computationally less demanding, capture much
of the physics of disc--planet interactions.
Thus they  may provide a useful
tool for studying the local interaction between forming
protoplanets and turbulent, protostellar discs.
\end{abstract}


\section{Introduction}\label{S0} 
The recent and ongoing discovery of extrasolar giant planets has stimulated
much current  investigation of theories of planet formation
(e.g. Mayor \& Queloz 1995;
Marcy, Cochran, \& Mayor 1999; Vogt et al. 2002). At the present time
giant planets are believed to originate
in protostellar discs. Possible formation scenarios are either through
direct gravitational
fragmentation of a young protostellar disc (e.g. Boss 2001) or through 
the accumulation of a solid  core
  which then undergoes   mass build up through gas accretion once the core mass
reaches $\simeq 15$ Earth masses (e.g. Bodenheimer \& Pollack 1986;
Pollack et al. 1996). In either scenario giant planets are likely
to form at significantly larger radii than those observed,
requiring a migration mechanism to bring them closer to the central star.
Disc--protoplanet interaction provides a natural mechanism.

In the standard picture a protoplanet exerts torques on a protostellar disc
through the excitation of spiral density waves 
(e.g. Goldreich \& Tremaine 1979).
These waves carry  an associated angular momentum flux
which is deposited in the disc 
where the
waves are damped. This process results in a negative  torque acting on the
protoplanet from the outer disc and a positive torque 
acting on it from the disc interior to its orbit, with the outer
disc torque normally dominating. For low mass
planets that induce linear perturbations in the disc, the resulting
migration is usually referred to as type I migration.

Previous  work on the interaction between a protoplanet and a  laminar
disc with Navier Stokes viscosity
(Papaloizou \& Lin 1984; Lin \& Papaloizou 1993) indicates that a
sufficiently massive protoplanet can open up an annular gap in the disc
centred on its orbital radius.
For  typical protostellar disc models the protoplanet
needs to be approximately a Jovian
mass for gap formation to occur.
Recent simulations (Bryden et al. 1999; Kley 1999; Lubow, Seibert, \&
Artymowicz 1999;
D'Angelo, Henning, \& Kley 2002)
examined the formation of gaps by giant protoplanets,
and also estimated the maximal gas accretion rate onto them.
The orbital evolution
of a Jovian mass protoplanet embedded in a standard laminar 
viscous protostellar
disc model
was studied by Nelson et al. (2000). They found that
that gap formation and accretion of the inner disc by the central mass
led to the formation of a low density inner cavity in which the planet
orbits.  Interaction with the outer disc resulted in inward
migration on a time scale of a few $\times 10^5$ yr. Here
the viscous evolution of the disc is responsible for
driving the inward migration of the giant planet. This mode of migration
is normally
referred to as type II migration.
Recent work examining gap formation,
mass accretion, and migration torques in three dimensions has been
presented by D'Angelo, Kley, \& Henning (2003) and Bate et al. (2003).
The disc models in these
studies all  adopted an anomalous
disc viscosity   modelled through the Navier--Stokes equations
without consideration of its origin.

The most likely origin of the viscosity is through
MHD turbulence resulting from the magnetorotational
instability (MRI) (Balbus \& Hawley 1991)
and it has recently become possible through
improvements in computational resources 
to  simulate  discs in which this  underlying mechanism responsible
for angular momentum transport is explicitly calculated.
This is necessary because the turbulent fluctuations 
do not necessarily result in transport phenomena that can be
modelled with the Navier Stokes equation.

To this end Papaloizou \& Nelson (2003) and Nelson
\& Papaloizou (2003a) (hereafter papers I and II) developed models of turbulent 
protostellar accretion discs and considered the interaction
with a giant protoplanet of $5$ Jupiter masses.
The large mass was chosen to increase the scale of the interaction
so reducing the computational resources required.
This protoplanet was massive enough to maintain 
a deep gap separating the inner and outer disc and 
exert torques characteristic of type II migration.
A study of gap formation in turbulent discs has also been presented by Winters, Balbus, \& Hawley (2003a)

It is the purpose of this paper to extend our previous 
work to a wider range of protoplanetary masses and disc parameters.
In particular we study the nature of the disc protoplanet
interaction as the protoplanet mass increases and investigate
the condition for gap formation.

To do this we utilise both global and local simulations.
The global simulations are more realistic but too computationally
demanding for significant  extended parameter surveys. On the other hand
local shearing box simulations being less computationally demanding
enable a wider parameter survey at higher resolution.
Of course because of the high degree of symmetry they
do not allow migration rates to be estimated but most other features
of the global simulations are reproduced.

In this paper we focus on the morphology of the disc response
as a function of the protoplanet mass leaving the analysis
of the disc--protoplanet torques resulting from the interaction
to a companion paper 
(Nelson \& Papaloizou 2003b -- hereafter paper IV).

\noindent The plan of this paper is as follows: \\
In \S \ref{S1} we describe both the 
initial global disc models and the shearing box models used for the simulations.

\noindent In \S \ref{S2} we discuss the nature of the waves that can propagate
and how their excitation leads to the existence of the prominent
wake that is frequently seen in the simulations of protoplanet disc interaction
even when the disc is turbulent.

\noindent In \S \ref{S3} we describe the numerical procedure used.

\noindent In \S \ref{S4} we present the results of the local and global
simulations.
Finally, we summarize our results in \S \ref{S5} .

\section{Set up of Initial Models } \label{S1} 

In this paper we have performed both global and local simulations
of protoplanets interacting with  model discs in which
MHD turbulence driven by the MRI operates. We consider simulations 
for which the magnetic field has zero net flux and therefore
an internally generated dynamo is maintained.
We determine the disc response as a function of the protoplanet mass
and as far as is possible  study its characteristics for the
different kinds of model focusing on the conditions
that determine whether the response is  weak (linear) or strong (non linear).
We now describe the model set up for the global and local simulations. 

\subsection{Global Simulations} \label{global-setup}
\noindent
The governing equations for MHD written in a frame rotating
with arbitrary  uniform  angular velocity $\Omega_p {\bf {\hat k}} $
with $ {\bf {\hat k}} $ being the unit vector along the rotation axis
assumed to be  in the vertical $(z)$  direction
are the continuity equation:
\begin{equation}
\frac{\partial \rho}{\partial t}+ \nabla \cdot {\rho\bf v}=0, \label{cont}
\end{equation}
the equation of motion
\begin{equation}
\frac{\partial {\bf v}}{\partial t}
 + {\bf v}\cdot\nabla{\bf v} + 2\Omega_p {\bf {\hat k}}{\bf \times}{\bf v} =
-{\nabla p \over \rho}  -\nabla\Phi +
{(\nabla \times {\bf B}) \times {\bf B}\over 4\pi\rho}, \label{mot}
\end{equation}
and the induction equation
\begin{equation}
\frac{\partial {\bf B}}{\partial t}=\nabla \times ({\bf v} \times {\bf B}).
\label{induct}
\end{equation}
where ${\bf v}, P, \rho, {\bf B}$ and $\Phi$ denote the fluid
velocity, pressure, density  and  magnetic field respectively.
The potential $\Phi = \Phi_{rot} +\Phi_G$
contains contributions due to gravity, $\Phi_G$
and the centrifugal potential $\Phi_{rot} = 
-(1/2)\Omega_p^2|{\bf {\hat k}}\times
{\bf r}|^2.$

The gravitational contribution is taken to be  due to a central mass  $M_*$
and planet with mass $M_p.$
Thus in cylindrical coordinates $(r, \phi, z)$ 
with the planet located at $(r_p, \phi_p, 0)$ and the star located at
$(r_*, \phi_*, 0)$
$\Phi_G = \Phi_c + \Phi_p ,$
where
\be \Phi_c = -{GM_* \over |{\bf r} - {\bf r_*} |} \ \  {\rm and} \ \
 \Phi_p = -{GM_p\over \sqrt{r^2 + r_p^2 -2rr_p\cos(\phi-\phi_p)+ b^2}}.\ee

\noindent
Here, as in papers I and II, for reasons of computational expediency,
we  have neglected the dependence
of the gravitational potentials on $z$
along with the vertical stratification of the disc.
Thus the global simulations are of cylindrical discs
(e.g. Armitage 1998, 2001; Hawley 2000, 2001; Steinacker \& Papaloizou 2002).
To model the effects of the   
reduction of the planet potential with
vertical height,  we have incorporated a softening length $b$ in  its
potential.

\noindent We  use a locally isothermal equation of state in the form
\begin{equation}
P=\rho\cdot c( r)^2,
\end{equation}
with $c$ denoting the sound speed
which is specified as a fixed function of the cylindrical
radius $r.$
The above gives the basic equations used for the global simulations.

\subsubsection{Initial and Boundary Conditions}
The global simulations presented here all use the same underlying 
turbulent disc model. This model has a constant aspect ratio 
$H/r \equiv c(r)/(r \Omega)=0.07$, where $\Omega$ is the disc angular
velocity as measured in the inertial frame. The inner radial boundary of
the computational
domain is at $R_{in}=1$ and the outer boundary is at $R_{out}=8$.
The boundary conditions employed are very similar to those described
in paper I. Regions of the disc in the close vicinity of the inner
and outer boundaries were given non--Keplerian angular velocity profiles
that are stable to the MRI, and which have large values of the
density in order to maintain radial hydrostatic equilibrium. These regions
act as buffer zones that prevent the penetration of magnetic field
to the radial boundaries, thus maintaining zero-net flux in the 
computational domain. The inner buffer zone runs from $r=1.14$ to $r=R_{in}$,
and the outer buffer zone runs from $r=7.0$ to $r=R_{out}$.

The disc was initiated with a zero net flux
toroidal magnetic field in a finite annulus in the disc between the 
radii $2 \le r \le 6$. The initial field varied sinusoidally over
four complete cycles within this radial domain, and the initial ratio of
the volume integrated magnetic energy to the volume integrated
pressure was $\langle 1/\beta \rangle=0.032$.
The initial disc model had azimuthal domain
running between $\phi=0$ to $\pi/4$, and vertical domain from $z_{min}=-0.21$ to
$z_{max}=0.21$. Periodic boundary conditions were employed in the vertical
and azimuthal directions.

The unperturbed disc model was evolved until the MRI produced a turbulent disc
model with a statistically steady turbulence. This model had a volume averaged
value of the ratio of magnetic energy to mean pressure of $\simeq 0.011$
and a volume averaged value of the Shakura--Sunyaev stress parameter
$\alpha \simeq 7 \times 10^{-3}$. In order to generate the full $2 \pi$ disc
models used to study the interaction between turbulent discs
and protoplanets, eight identical copies of this relaxed disc sector 
were joined 
together.
The model described in table~\ref{table2} with azimuthal domain $\pi/2$ (run G5)
contained
only two copies of the $\pi/4$ sector, and periodic boundary conditions were 
employed in the azimuthal domain.

Gravitational softening of the protoplanet was employed in all
simulations. In runs G1--G4 described in table~\ref{table2}, the softening
parameter $b=0.33H$. For run G5, this was reduced to $b=0.1H$.
For runs G1, G2, and G3, the planet was placed at a radius such that 
$|r_p-r_*|=3$
and at azimuthal location $\phi_p=\pi$
in the disc. For runs G4 it was placed such that $(|r_p-r_*|,\phi_p)=(2.5,\pi)$
and for run G5 it was placed at $(|r_p-r_*|,\phi_p)=(2.5,\pi/4)$.
All simulations were performed in a rotating frame whose
angular velocity equalled that of the circular Keplerian planetary orbit.
The unit of time adopted when discussing the results is
$\Omega^{-1}$ evaluated at $r=1$, the inner boundary of
the computational domain. We note that the orbital period at this
radius is $P(r=1)=2\pi$, the orbital period of a planet at
$r_p=2.5$ is $P(r=2.5)\simeq 24$, and the orbital period of a planet
at $r_p=3$ is $P(r=3)\simeq 32$.

In order to explore the disc--planet interaction over a wider range
of parameter space and with higher resolution in the vicinity of the planet,
we have adopted a local shearing box approximation. This we describe below.

\subsection{Local Shearing Box Simulations}

In the shearing box limit (Goldreich \& Lynden-Bell 1965)
we consider a small  Cartesian box centred on the point at which the Keplerian
angular velocity is $\Omega_p.$
We define local Cartesian coordinates $(x,y,z)$ with associated
unit vectors $({\bf i}, {\bf j}, {\bf k}).$ The direction ${\bf i}$
is taken to be outward along the line joining the origin
to the central object while ${\bf k}$ points in the vertical direction.
The direction ${\bf j}$ is along the unperturbed rotational  shear flow.

The equation of motion may then be written as

\begin{equation}
\frac{\partial {\bf v}}{\partial t}
+{\bf v}\cdot\nabla{\bf v} 
+ 2\Omega_p {\bf {\hat k}}{\bf \times}{\bf v} -3\Omega_p^2x{\bf i}=
-{\nabla p \over \rho} -\nabla\Phi_p +
{(\nabla \times {\bf B}) \times {\bf B} \over 4\pi \rho}. \label{boxmot}
\end{equation}

The induction and continuity equations retain the same form
as equations (\ref{induct}) and (\ref{cont}) but are of course
expressed in component form in the local box centred
Cartesian coordinates.
We recall that the term $\propto x$ in equation (\ref{boxmot})
is derived from a first order Taylor expansion  about $x=0$
of the combination
of the 
gravitational  acceleration due to the central mass and the centrifugal
acceleration (Goldreich \& Lynden-Bell 1965).
However, as for the global disc simulations, we have neglected the $z$ dependence
of the gravitational potential and therefore  vertical
stratification. Thus in this respect the simulations are of unstratified
boxes of the type considered by Hawley, Gammie \& Balbus (1995).

Here, when introduced, the planet is located in the centre 
of the box, thus

 \be \Phi_p = -{GM_p\over \sqrt{x^2+ y^2+ b^2}}.\ee

In the steady state  box with no planet or magnetic field, the equilibrium
velocity is due to Keplerian shear
and thus 
\be {\bf v} = (0, -3\Omega_p x/2 ,0). \label{boxv}\ee

We adopt an isothermal equation of state $P = \rho c^2,$ with $c$
being the sound speed which is taken to be constant throughout the box.

\subsubsection{Initial and Boundary Conditions}

The shearing box is presumed to represent a local patch of a differentially
rotating disc. Thus the appropriate boundary conditions on the bounding  faces
$y = {\rm constant} = \pm Y$ and $z = {\rm constant} = \pm  Z$ derive
from the requirement of periodicity in the local Cartesian coordinate
directions normal to the boundaries.
On the boundary faces $x = {\rm constant} = \pm X,$ 
the boundary  requirement is for periodicity in local shearing coordinates.
Thus for any state variable
$F(x,y,z),$ the condition is that $F(X,y,z) = F(-X, y -3\Omega_p X t ,z).$
This means that information on one radial boundary face is communicated
to the other boundary face at a location in the azimuthal coordinate
$y$  shifted by  the distance
the faces have sheared apart since the start of the simulation, $t,$
(see Hawley, Gammie \& Balbus 1995).

For all models adopting a shearing box,
the softening parameter was taken as $b=0.3H.$
The potential was flattened (made to attain a constant value
in a continuous manner)
at large distances from the protoplanet in order to satisfy the
periodic boundary conditions.
In simulations Ba0--Ba4 (described in table~\ref{table1})
this was done outside the cylinder of radius $3H$
centred on the planet. In simulations Bb1--Bb4 this was done only close
to the boundary.

The simulations  Ba0--Ba4 occupied a total width $8H$ in $x$
and $4\pi H$ in $y.$ These boxes were found to be large
enough to accomodate the scale of the disc response to forcing by the
protoplanet and enable gap formation.
But note that interference from  protoplanets in neighbouring
boxes implied by the periodic boundary conditions   is
 not eliminated entirely. 
In the   small protoplanet mass 
regime waves propagate between boxes and in the nonlinear regime
torques exerted by protoplanets in neighbouring boxes may influence
the gap formation process. 
However, comparison between local and global runs indicates that such effects
are not major and do not prevent the box runs from capturing
the essential physics.

To examine the effect of varying box size we ran simulations
Bb1--Bb4 which were of the same resolution but had twice
the extent in $y.$

All of these models were initiated by imposing a vertical field,
with zero net flux through the box,
varying  sinusoidally in $x$ with a wavelength of $H.$
The amplitude was such that the initial value of the ratio
of the total magnetic energy to volume integrated pressure
was $0.0025.$ The  initial velocity in the $x$ direction at each
grid point  was
chosen to be the product of  a random number between $-1.0$ and $1.0$
and $0.1c.$
The initial velocity in the $y$--direction is given by equation~\ref{boxv},
and the $z$ component of velocity was initialized to zero.
The unit of time for all simulations was taken to be $\Omega_p^{-1}.$
In these units the orbital period at the centre of the box is
$2\pi.$

 \begin{table*}
 \begin{center}
 \begin{tabular}{|l|l|l|l|l|l|l|l|l|l|l|}\hline\hline
       &     &     &     &     &        &        &        &          \\
 Model&$z$ domain&$x$ domain&$y$ domain &$t_1$&$t_2$&$(M_p R^3/(M_* H^3)$&$n_z$&$n_x$&$n_y$\\
       &     &     &     &     &        &        &        &     &     &           \\
 \hline
 \hline
 Ba0&$(-H/2, H/2)$& $(-4H, 4H)$ & $(-2\pi H, 2\pi H)$ & $0.0$&  $637$&$0.0$&$35$&$261$&$200$\\
 Ba1&$(-H/2, H/2)$& $(-4H, 4H)$ & $(-2\pi H, 2\pi H)$ & $354$&  $621$&$0.1$&$35$&$261$&$200$\\
 Ba2&$(-H/2, H/2)$& $(-4H, 4H)$ & $(-2\pi H, 2\pi H)$ & $354$&  $644$&$0.3$&$35$&$261$&$200$\\
 Ba3&$(-H/2, H/2)$& $(-4H, 4H)$ & $(-2\pi H, 2\pi H)$ & $354$&  $619$&$1.0$&$35$&$261$&$200$\\
 Ba4&$(-H/2, H/2)$& $(-4H, 4H)$ & $(-2\pi H, 2\pi H)$ & $354$&  $650$&$2.0$&$35$&$261$&$200$\\
 Bb1&$(-H/2, H/2)$& $(-4H, 4H)$ & $(-4\pi H, 4\pi H)$ & $216$&  $392$&$0.2$&$35$&$261$&$396$\\
 Bb2&$(-H/2, H/2)$& $(-4H, 4H)$ & $(-4\pi H, 4\pi H)$ & $216$&  $431$&$0.5$&$35$&$261$&$396$\\
 Bb3&$(-H/2, H/2)$& $(-4H, 4H)$ & $(-4\pi H, 4\pi H)$ & $216$&  $411$&$1.0$&$35$&$261$&$396$\\
 Bb4&$(-H/2, H/2)$& $(-4H, 4H)$ & $(-4\pi H, 4\pi H)$ & $216$&  $485$&$2.0$&$35$&$261$&$396$\\
\hline
\end{tabular}
\end{center}
\caption{ \label{table1}
Parameters of the shearing box simulations:
The first column gives the  simulation label, the second,  third and fourth
give the extent of the coordinate domains considered.  
The first nine simulations were performed
in shearing boxes  with the $x,y,$ and $z$ domains referring to the Cartesian domains.
The last two simulations
were global   so that in these cases the  
$x,y,$ and $z$ domains refer to the $r,\phi ,$ and $z$ domains  for the cylindrical
coordinate system used. 
The  fifth  and sixth column
give  the start and end times if the simulation measured in dimensionless units.
Simulations with protoplanets were started by inserting the protoplanet into  the turbulent
model generated from the simulation Ba0 at time indicated. The seventh column
gives the value of $(GM_p/(\Omega^2 H^3).$ The eighth, ninth and tenth columns
give the number of computational grid points in the $x,y,$ and $z$ coordinates respectively.
These numbers include any ghost zones used to handle boundary conditions.} 
\end{table*}

\subsection{Dimensionless Variables and Scaling with Planet Mass}\label{Scaling}
It is helpful to write the governing equations for the shearing box
in dimensionless form. In this way the dependence of the 
outcomes of the simulations on 
planet mass and disc parameters is made clearer (see also Korycansky \&
Papaloizou 1996). To do this we
adopt dimensionless coordinates $(x',y',z') = (x/H,y/H,z/H),$
where the length scale $H = c/\Omega_p$ would correspond to the disc scale height
were it vertically  stratified and the dimensionless time $t' = \Omega_p t.$
We also note that the Keplerian relation
$\Omega_p^2 = GM_*/R^3,$ may be used to relate the rotation rate
of the centre of the box to a putative central mass $M_*$ and orbital
radius $R.$
We now introduce the dimensionless velocity, sound
speed and planet mass through
${\bf v}' = {\bf v}/(\Omega_p H),$ $c' = c/(\Omega_p H),$ and
$M_p' = M_p R^3/(M_* H^3)$ respectively.
As we do not include  the self-gravity of the disc in calculating
its response, the magnitude of the disc density may be arbitrarily
scaled. The models simulated here all start with a uniform density
$\rho_0$ which, together with $H,$ may be used to specify the mass scaling.
Then  we define a dimensionless  density and magnetic field through
$\rho' = \rho_0 \rho$ and  
${\bf B}' =  {\bf B}/(\Omega_p H \sqrt{4\pi\rho_0}).$
The equation of motion may be written in the
above dimensionless variables in the form

$$
\frac{\partial {\bf v}'}{\partial t'}
+{\bf v}'\cdot\nabla{\bf v}'
+ 2{\bf {\hat k}}{\bf \times}{\bf v}' -3x'{\bf i}= \ \ \ \ \ \ \ \ \ \ \ \ $$
\begin{equation} -{\nabla (\rho' c^{'2}) \over \rho'} -\nabla\Phi_p' +
{(\nabla \times {\bf B}') \times {\bf B}' \over \rho'}, \label{boxmotdim}
\end{equation}
where
\be  \Phi_p' = -{M_p' \over \sqrt{x^{'2} + y^{'2}  + b^{'2}}},\ee
with dimensionless softening parameter  $b' = b/H$
and of course the spatial derivatives are with respect to the dimensionless coordinates.
Similarly the dimensionless induction equation becomes
\begin{equation}
\frac{\partial {\bf B}'}{\partial t'}=\nabla \times ({\bf v}' \times {\bf B}').
\label{inductdim}
\end{equation}
and the dimensionless continuity equation is
\begin{equation}
\frac{\partial \rho'}{\partial t'}+ \nabla (\cdot \rho' {\bf v}') = 0. \label{contdim}
\end{equation}

We remark that in dimensionless coordinates
the boundaries of the box are $x' =\pm X/H,$  $y' =\pm Y/H,$ $z' =\pm Z/H.$
Also if the magnetic field has zero net flux and is dynamo generated and
thus a spontaneous product of the simulation, given that
 the only parameter occurring in equations (\ref{boxmotdim} - 
\ref{contdim} ) is the dimensionless mass $M_p' = M_p R^3/(M_* H^3),$
we should be able to  consider the dependence of the 
time averaged outcome of a simulation to be only on 
$ X/H, Y/H, Z/H,$
and $M_p' = M_p R^3/(M_* H^3).$

For a planet embedded in a box with natural vertical scale  $H$
with sufficiently large $X/H$ to incorporate the tidal interaction
without interference from other boxes centred on different radii,
the only parameters determining long term averaged properties of a simulation
should be
$Y/H$
and $M_p'.$ Thus for such a quantity $Q,$ we should have
$Q = f(Y/H, M_p R^3/(M_* H^3)).$

Note that the above discussion implies that for a box of fixed dimension, the
only distinguishing parameter is $ M_p' =  M_p R^3/(M_* H^3).$ 
This parameter may also be interpreted as the cube of the  ratio of the Hill
radius to disc scale height and the condition that  $M_p' \gs  1$ 
leads to the thermal condition of 
Lin \& Papaloizou (1993) that the Hill radius should exceed the disc
scale height for gap formation to occur.   From Korycansky \& Papaloizou (1996) 
this condition is also required in order that the perturbation
due to the protoplanet be non linear.

The above discussion indicates that were we to vary the box size in the azimuthal
$(y)$ direction, the condition
for gap formation should be of the form $M_p' > f_g( Y/H).$
As gap formation by the same planet in a longer box is expected 
to be more difficult,
we expect $f_g$ to be a non decreasing function of its argument.
In order to correspond to a full circle of radius $R,$ we require
that $Y = \pi R.$

It is perhaps suggestive to point out that if
$f_g(\pi R/H)
= 40\alpha (R/H),$ the condition for gap formation indicated above becomes
$ M_p /M_* > 40\alpha H^2/(R^2)$
which is of the same form as the viscous criterion of
Lin \& Papaloizou (1993) which
was obtained by requiring angular momentum transport
due to excited waves to exceed that due to viscosity.
However, it is important to note that if we introduce such an 
$\alpha$ parameter  here, it cannot correspond to an imposed viscosity
as in Lin \& Papaloizou (1993). Instead it must be related
to the turbulent transport arising from MHD turbulence.
We take  it to correspond to a time average of the volume averaged stress  
$\langle B_xB_y/(4\pi) \rangle $ expressed in units of the volume
averaged pressure $\langle P \rangle,$ thus corresponding to the 
Shakura \& Sunyaev (1973) parameterization. This procedure is reasonably
well defined in the parts of the disc not strongly perturbed by a protoplanet
(eg.  Paper II). For simulations with zero net magnetic
flux $\alpha \sim 5\times 10^{-3}$ but it takes on larger values
when a net poloidal or toroidal flux is imposed
(eg. Hawley, Gammie \& Balbus 1996; Steinacker \& Papaloizou 2002). 

Adopting  $\alpha  = 5\times 10^{-3},$
suggests $f_g(Y/H)
\sim 0.07(Y/H).$ We recall that should $0.07(Y/H)$ be less than unity,
the requirement of nonlinear perturbation by the protoplanet in order
for gap formation to occur is $ M_p /M_* >  C_t(H/R)^3,$  
where $C_t$ is a constant of order unity (Korycansky \& Papaloizou 1996).
Both these conditions can be combined to give the requirement that
\begin{equation}
M_p R^3/(M_* H^3) > \rm{ max}  (C_t, 0.07(Y/H)) \label{condres}
\end{equation}
In this way we may see how both the thermal and viscous conditions
of Lin \& Papaloizou (1993) may be related to the simulations.

\section{Wakes and Characteristics} \label{S2}
A well known phenomenon found in simulations of
protoplanets interacting with
laminar discs
is a prominent wake that is observed to trail away from the protoplanet
eventually becoming a spiral wave propagating away from it
(e.g. Kley 1999; Nelson et al 2000; Lubow, Seibert \& Artymowicz 1999; Ogilvie \& Lubow 2002).
Such  structures, though distorted,  are also seen in our simulations of
discs with MHD turbulence in which there are strong and highly
time variable fluctuations, even for weakly perturbing protoplanets.
In view of their importance and persistence we here give a brief
discussion of their interpretation as being defined by 
a characteristic  ray emanating from the protoplanet that separates the flow
shearing past it into parts causally connected and not causally
connected by waves. The wakes thus  naturally correspond to the location
a strong wave 
sent out from the protoplanet into the flow shearing past it.  

\begin{table*}
 \begin{center}
 \begin{tabular}{|l|l|l|l|l|l|l|l|}\hline\hline
       &     &     &     &     &        &        &              \\
 Model&$\phi$ domain&$H/r$&$M_p/M_*$& $(M_p R^3/(M_* H^3)$&$n_r$&$n_{\phi}$&$n_z$\\
       &     &     &     &     &        &        &        
  \\
 \hline
 \hline
G1 & $2 \pi$ & 0.07 & $1 \times 10^{-5}$ & 0.03 & 450 & 1092 & 40 \\
G2 & $2 \pi$ & 0.07 & $3 \times 10^{-5}$ & 0.09 & 450 & 1092 & 40 \\
G3 & $2 \pi$ & 0.07 & $1 \times 10^{-4}$ & 0.30 & 450 & 1092 & 40 \\
G4 & $2 \pi$ & 0.07 & $3 \times 10^{-3}$ & 8.75 & 450 & 1092 & 40 \\
G5 & $\pi/2$ & 0.07 & $3 \times 10^{-3}$ & 8.75 & 450 & 276 & 40 \\
\hline
\end{tabular}
\end{center}
\caption{ \label{table2}
Parameters of the global simulations: The first column gives the simulation
label, the second gives the extent of the azimuthal domain, the third gives
the $H/r$ value of the disc, and the fourth gives the
protoplanet-star mas ratio. The fifth gives the ratio of $M_p/M_*$ to
$(H/r)^3$. The sixth, seventh, and eighth columns describe the number of
grid cells used in each coordinate direction.}
\end{table*}

\subsection{The Characteristic Rays}
The set of basic equations defined by (\ref{cont}), (\ref{mot})  and (\ref{induct})
are well known to form a hyperbolic set which can be written in the standard form
(Courant \& Hilbert 1962)
\be  \sum_{i,j} \left( A_{ij}^{(k)}{\partial U_i \over \partial x_j} \right)
 + B^{(k)} =0  \ \ \ \ ( k =1, 2, ....,n) .\label{CH} \ee
Here the set of equations (\ref{cont}), (\ref{mot})  and (\ref{induct})
are arranged into a set of $n$  simultaneous first order  PDEs for the $n$
components $U_i$ of ${\bf U}$ which are in turn composed of $\rho$
together with the components of  ${\bf B}$
and ${\bf v}.$ The $x_j$ $(j = 1 - 4)$  are the three 
spatial coordinates and the time  $t$
respectively.
The characteristic rays, applicable to both the 
global and local formulation may be found by first obtaining the 
local dispersion relation associated with (\ref{CH}).
To do this one sets $U_i = U_{i0}\exp(i k _j x_j -i\omega t)$
under the assumption that the magnitudes of the  components of the wave vector
$k_l ( l =1, 2, 3)
\equiv {\bf k} =( k_x, k_y, k_z) $ and the  wave frequency  $|\omega|$ are very large.
The dispersion relation is then given by
the vanishing of a determinant of coefficients
$|A_{ij}^{(k)}k_j - A_{i4}^{(k)}\omega | =0.$
As is very well known (eg. Campbell 1997)  this dispersion relation
leads to three  distinct wave modes. The first are the
  Alfv{\'e}n waves propagating in opposite directions  which have 
$\omega = - {\bf k}\cdot {\bf v} 
\pm {\bf k}\cdot {\bf v}_A,$ with  $ {\bf v}_A =v_A {\bf B}/|{\bf B}|,$
where $v_A$ is the Alfv{\'e}n speed being such that
$v_A^2 = |{\bf B}|^2/(4 \pi \rho).$
The other modes are fast and slow  magnetosonic waves which
satisfy
\be ( \omega + {\bf k}\cdot {\bf v})^2 =
{k^2(c^2 + v_A^2) \over 2} \pm {\sqrt{ k^4(c^4 + v_A^4)
-2k^2 c^2 ({\bf k}\cdot{\bf v}_A)^2)} \over 2}.\ee
Here $k = |{\bf k}|$ and 
we note that the frequency, $\omega$  occurs added to the quantity
${\bf k}\cdot {\bf v}$ in these relations. This is due to
the Doppler shift occurring on account of the motion of the fluid.

For each of these waves we may specify $\omega =\omega({\bf k},{\bf r},t).$
Then the characteristic rays follow from integrating the equations
of geometric optics (e.g. Whitham 1999)
\be {d{\bf r}\over dt} =  {\partial {\omega}\over \partial {\bf k}} , \ \ \ {\rm and} \ \ \
{d{\bf k}\over  dt} = - {\partial {\omega}\over \partial {\bf r}}. \label{chare}\ee

To be specific, 
we here concentrate on the fast magnetosonic mode which in our case
leads to the most rapid communication. Further, as 
for the most part the magnetic energy is on the order of one percent
of the thermal energy, we shall neglect the field so that the mode
becomes an ordinary sound wave such that
\be  \omega = -  {\bf k}\cdot {\bf v} + ck. \ee

We consider the general situation to be one for which
the flow streams past the protoplanet from $y > 0$ to $y < 0$ for $x >0$
and with the flow moving in the opposite sense for $x < 0.$
Considering without loss of generality the case $ y > 0,$
when the flow streams past supersonically,  if we neglect
dependence on $z,$ it is divided into two
parts   by the characteristic originating
close to the planet from the point with  $y=0$ and the smallest possible
positive value of $x$ at which the flow speed is equal to the wave speed.
 If the planet had no gravity,
the fluid that is upstream of this characteristic
would be unaware of the presence of the planet.

The fluid that first knows about the planet, namely that on the characteristic
is a natural channel for waves to propagate
that are excited by the planet which is why this region is
so prominent in the simulations. Note
that the concept of  characteristic rays
is robust in that it can apply to a general
time dependent fluid with turbulence. Also with $z$ dependence
restored we would expect the fluid to be divided by a surface made up
from many rays rather than a single ray which is of the same form independent
of $z.$ This would lead to some blurring of the wake.

A characteristic ray is defined by its point of origin (here specified above)
and the value of ${\bf k}.$ However, note that only the ratio of the components
and not the magnitude matters. The natural choice is $k_z =0,$ and
$k_y/k_x$ such that $\omega =0.$ This is because
we expect a zero frequency wave to be excited by the planet in a frame
in which that is at rest. However, note that in a turbulent
time dependent  situation $\omega$ would not be constant.

The simplest situation is one when the flow is time independent
and the velocity is given by (\ref{boxv}).
In that case it follows  from equations (\ref{chare}) that
both $\omega = 0$ and $k_y$ are constant.

It further follows  from equations (\ref{chare}) 
 that the characteristic  rays then satisfy

\be {dy \over dx} = -\sqrt{v_y^2/c^2 - 1}. \ee

Using equation ({\ref{boxv}) for $v_y$ and integrating
we obtain for the characteristic ray for $x > x_0$

\be y/x_0 = -{x\over 2x_0}\sqrt{x^2/x_0^2-1} +
 {1\over 2}ln\left(\sqrt{x^2/x_0^2-1} + x/x_0\right), \label{crcv}\ee
with $ x_0 = 2H/3.$
To complete the picture we remark that the characteristic for
$x<0,$ can be found from the   condition $y(-x) = -y(x).$

\begin{figure}
\centerline{
\epsfig{file= 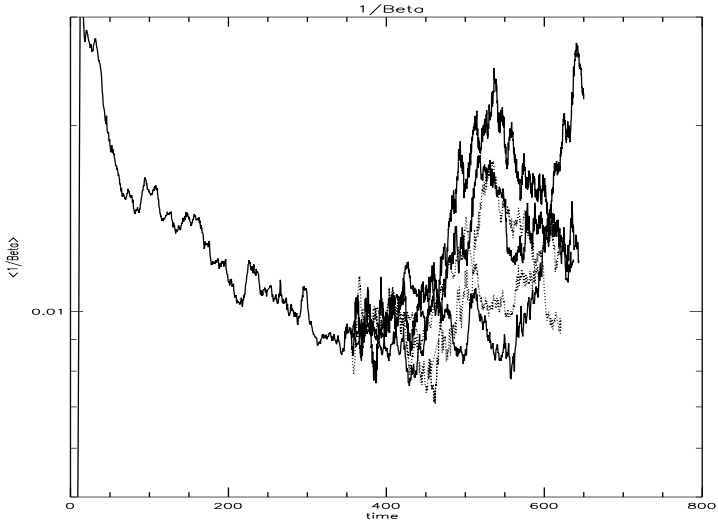 ,height=7.cm,width=10.cm,angle =0}}
\caption[]
{The ratio of the total magnetic energy to the volume integrated pressure,
$\langle 1/\beta \rangle,$ is plotted
as a function of time for models Ba0,  Ba1, Ba2, Ba3, and Ba4.
The plots begin  to separate after the protoplanet is inserted
at time $354.$. At time $500$ the curves correspond to simulations
Ba4, Ba1, Ba2,Ba3 and Ba0 with decreasing values of $\langle 1/\beta \rangle$
respectively. Although there are large fluctuations, the maximum
variation at any time is factor of $~3.$
This is significantly
smaller than the density depression in the deepest gap (see figure \ref{fig6}).}
\label{fig8}
\end{figure}

We comment that the computational domains in all of our simulations are
extensive enough to contain the characteristic associated with the waves
emitted by an exciting object, be it turbulent fluctuations or a protoplanet
(i.e. $x$ exceeds $2/3H$ by a significant margin).

\section{Numerical Procedure} \label{S3}
The numerical scheme that we employ is
based on a spatially second--order accurate method that computes the
advection using the monotonic transport algorithm (Van Leer 1977).
The MHD section of the code uses the
method of characteristics
constrained transport (MOCCT) as outlined in Hawley \& Stone (1995)
and implemented in the ZEUS code.
The code has been developed from a version
of NIRVANA originally written by U. Ziegler
(see SP, Ziegler \& R\"udiger (2000), and
references therein)

\begin{figure}
\centerline{
\epsfig{file= 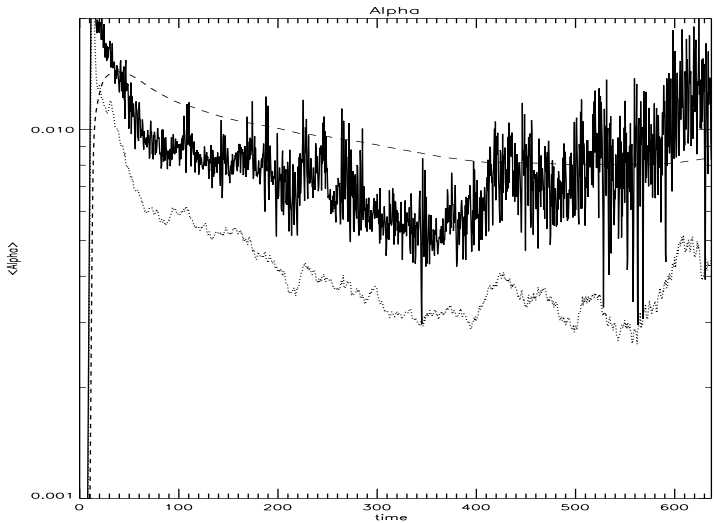 ,height=7.0cm,width=10.cm,angle =0} }
\caption[]{
 The  volume averaged   stress parameter
 $ \langle \alpha \rangle$ is plotted
as a function of time for model Ba0 which has no protoplanet.
The noisy full curve contains contributions from both the magnetic
and Reynolds' stresses while the dotted curve gives the magnetic
contribution. The dashed curve gives the running time average
of the total stress. The large initial values are due to
the initial strong instability which produces a channel phase.
However, after about four orbits at the centre of the box, the
running  time average is within a factor of two of the final value.
}
\label{fig9}
\end{figure}

\begin{figure}
\centerline{
\epsfig{file= 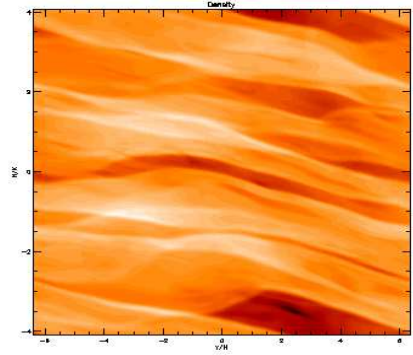,height=7.cm,width=5.cm,angle =0} }
\caption[]{  Density contour plot in a $(x,y$) plane for simulation
Ba0 which has no embedded protoplanet. The extensions
are $8H$ in the $x$ direction and $4\pi H$ in the $y$ direction.
The plot applies near the end of the simulation. The range of variation
in the density is about a factor of three.
This is typical of what is obtained in any $(x,y)$ plane once a statistically
steady state turbulence is set up. Note the trailing elongated structures.
}
\label{fig1}
\end{figure}

\noindent
\begin{figure}
\centerline{
\epsfig{file=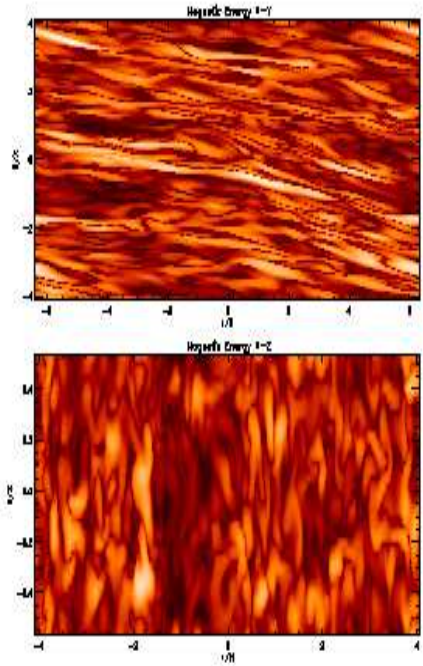,height=14.cm,width=7.cm,angle =0} }
\caption[]{  Contours of magnetic energy density corresponding to the plot
in figure \ref{fig1}. The upper plot corresponds to the $(x,y)$ plane
illustrated in figure \ref{fig1} while the lower plot is taken in a
characteristic $(x,z)$ plane. The extensions are $H$ in the $z$ direction
and $8H$ in the $x$ direction. Due to regions of very small magnetic field,
the range of variation
of magnetic energy density is much larger than that of  the  density
being as much as $\sim 10^7.$ These  plots are typical of
what is found  for the turbulence once it has reached a statistically steady state.
Note that trailing elongated features are also apparent in the
distribution of  the magnetic energy density.
}
\label{fig2}
\end{figure}

\subsection{Vertically and Horizontally Averaged Stresses
and Angular Momentum Transport}\label{S3a}
\noindent
In order to describe average properties of the turbulent models,   following papers I and II
we use quantities that are
vertically and azimuthally averaged over the $(y , z)$ domain for the shearing boxes
 or the $(\phi ,z )$ domain for the global simulations  (e.g. Hawley 2000).
Thus we may consider $y \equiv \phi$ and $x \equiv r$  in these cases.
Sometimes an additional time average may be adopted.
The vertical and  azimuthal average of some quantity $Q$ is  then defined through

\noindent
\begin{figure}
\centerline{
\epsfig{file= 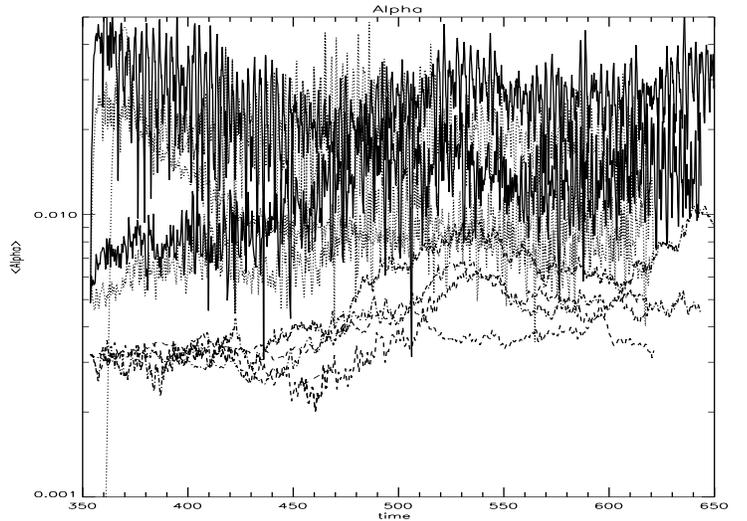 ,height=7.cm,width=10.cm,angle =0} }
\caption[]{
The  volume averaged  total  stress parameter
$\langle \alpha \rangle$ is plotted
as a function of time for models Ba1, Ba2, Ba3, and Ba4.
The plots begin  from the moment the protoplanet is inserted.
The noisy  upper curves contain contributions from both the magnetic
and Reynolds' stresses. They can be identified at early times
as for Ba4 (uppermost full curve), Ba3 (uppermost dotted curve),
Ba2 (next highest full curve), Ba1 (next highest dotted curve).
 The lower and much less noisy  curves  give the magnetic
contribution. These are similar until they separate at about time $500.$
At this time they may be identified  with increasing
values of the stress parameter as associated with
 Ba1,Ba3, Ba2 and  Ba4 respectively.
The  values  of  the total $\langle \alpha \rangle$ show a sharp increase
when the protoplanet is inserted in the cases with higher mass protoplanets
on account of the excitation of density waves which
provide angular momentum transport.
}
\label{fig10}
 \end{figure}

\begin{equation}
{\overline {Q}} ={\int \rho  Q dz dy \over \int  \rho dz dy}.
\end{equation}

\noindent The  disc surface density is given by
\begin{equation}
\Sigma = {1\over 2Y }\int \rho dz dy.
\end{equation}

\noindent  The  vertically and azimuthally
averaged   Maxwell and
Reynolds stresses,  are  given by:
\begin{equation}
T_M= 2Y
\Sigma{\overline{\left({B_x B_y \over 4\pi\rho}\right)}}
\end{equation}
and
\begin{equation}
T_{Re}=2Y
\Sigma
{\overline{\delta v_x\delta v_y}}.
\end{equation}
 Here the velocity fluctuations $\delta v_x$ and $\delta v_y$
are defined through,
\begin{equation}
\delta v_x=v_x-{\overline{v_x}},
\end{equation}
\begin{equation}
\delta v_y=v_y- {\overline{v_y}}.
\end{equation}
The horizontally and vertically averaged  Shakura  \& Sunyaev (1973)
$\alpha$ stress parameter appropriate to the
total stress is  given by
\begin{equation}
\alpha=\frac{T_{Re}-T_M}{2Y
\Sigma{\overline{ \left(P/\rho\right)}}},
\label{alpha}
\end{equation}
The angular momentum flow across a line of constant $x$
is given by
\begin{equation}
 {\cal F} =\Sigma W \alpha\overline{P/\rho}.
\label{transport}
\end{equation}
Here $W= R^2$ for the shearing box simulations  or $W= r^2$
for the global simulations.

\section{Numerical Results} \label{S4}
We now present numerical results for the evolution
of the simulations.

\begin{figure}
\centerline{
\epsfig{file=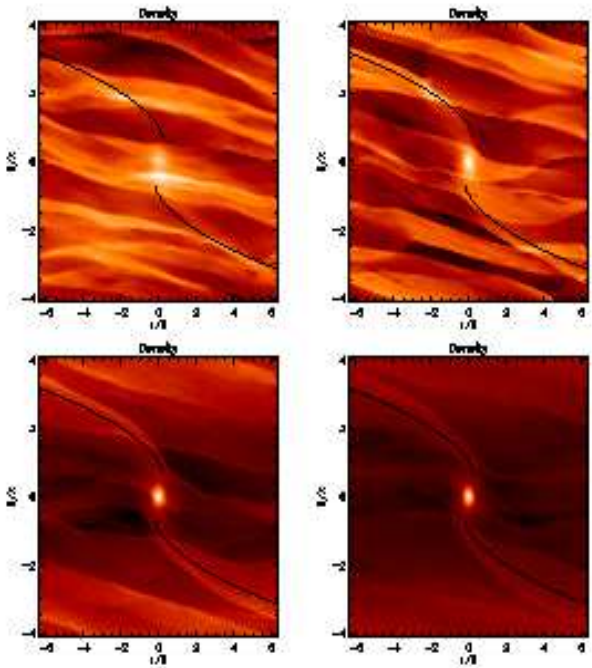,height=14.cm,width=10.cm,angle =0} }
\caption[]
{ Density contour plots in a typical $(x,y$) plane for simulations
Ba1, top left, Ba2 top right, Ba3 bottom left, and Ba4 bottom right
taken near the end of the simulations.
These had $M_p R^3/(M_* H^3) = 0.1, 0.3, 1.0$ and $2.0 $ respectively.
As $M_p R^3/(M_* H^3)$ increases, the wake becomes more prominent,
material is accreted by the protoplanet and is pushed toward the radial
boundaries as a gap is formed. The location of the   characteristic ray
given by equation (\ref{crcv}) is also plotted.
The qualitative structure seen in these simulations is maintained
once a quasi steady state has been attained.}
\label{fig3}
\end{figure}

\begin{figure}
\centerline{
\epsfig{file=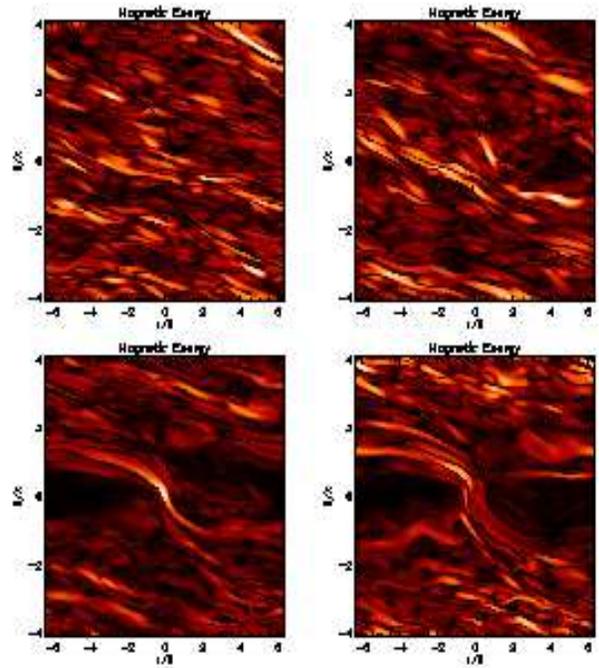,height=14.cm,width=10.cm,angle =0} }
\caption[]
{ As in figure \ref{fig3} but in this case magnetic energy  density contours are plotted.
 As $M_p R^3/(M_* H^3)$  increases and a gap forms, the  magnetic energy  generally
decreases in the gap region. However, some remains concentrated in regions
near the high density wakes. The qualitative structure seen in these simulations is maintained
once a quasi steady state has been attained.}
\label{fig3a}
\end{figure}
\subsection{Shearing Box Simulations}
All of the shearing boxes are derived from the simulation Ba0 which was carried out
without a protoplanet. It settled into a quasi steady state with conserved
zero net flux and  dynamo maintained 
turbulence as has  been been described by others (e.g. Hawley, Gammie \& Balbus 1996).
This model was run for $100$ orbits at the box centre.
The ratio of the total magnetic energy to the volume integrated pressure,
$ \int {\bf B}^2/(8\pi)dV/ \int PdV  = \langle 1/\beta \rangle,$ is plotted
as a function of time  in figure \ref{fig8}. 

After running for $354$ time units, this model was  used to provide initial
conditions for models with  protoplanets of varying masses.
Simulations  Ba1, Ba2, Ba3, and Ba4 were begun in this way 
with $M_p R^3/(M_* H^3) = 0.1, 0.3, 1.0$ and $2.0 $ respectively.
We also  plot 
$ \int {\bf B}^2/(8\pi)dV/ \int PdV  = \langle 1/\beta \rangle$ 
as a function of time  in figure \ref{fig8}  for models  Ba1, Ba2, Ba3, and Ba4.
The plots begin  to separate from each other
shortly after  after the simulations  with  protoplanets
have begun. Note that although $\langle 1/\beta \rangle$   is in
general confined between $0.008$
and $0.02$ with or without a protoplanet, there are strong fluctuations
on a range of timescales down to the orbital timescale.
The existence of such fluctuations and the difficulties they 
cause for defining mean state variables for the turbulent flow
has been emphasised recently by several authors 
(e.g. papers I and II; Winters, Balbus \& Hawley 2003b; Armitage 2002).

The volume averaged  stress parameter
$\langle \alpha \rangle$ is plotted
as a function of time for model Ba0 in figure \ref{fig9}.
As for $\langle 1/\beta \rangle, $ erratic fluctuations are seen.
The sum of the magnetic
and Reynolds' stresses   nearly always exceeds the magnetic
contribution for which the short term fluctuations are significantly
less severe. In these respects the behaviour
of local and global simulations are similar (Steinacker \& Papaloizou 2002,
papers I and II, and section~\ref{global-res}).
The running time average of $\langle \alpha \rangle$ is also
plotted and this shows much smoother behaviour.
Somewhat large initial values  occur because of 
the initial strong instability which produced a channel phase. However,
after about four orbits at the centre of the box, the
running  time average is within a factor of two of  its final value of $0.008$.
This value is in agreement with that given in 
Winters, Balbus \& Hawley (2003b).
But note that the running time average still shows variations at the $10$
percent level after $100$ orbits.

The pattern of turbulence in simulation Ba0 is established after several orbits.
A typical density contour plot in a $(x,y$) plane 
is presented in figure \ref{fig1}
near the end of the simulation. The range of variation
in the density in this plane is about a factor of three.
The situation illustrated  here
is typical of what is obtained in any $(x,y)$ plane once a statistically
steady state turbulence is set up. Elongated trailing structures are noticeable.
We plot contours of the magnetic energy density in figure~\ref{fig2}
which correspond to the plot
in figure \ref{fig1}. The upper plot corresponds to the $(x,y)$ plane
illustrated in figure~\ref{fig1} while the lower plot is taken in a
characteristic $(x,z)$ plane. Due to the existence of regions where
the magnetic field is near zero, the range   
of the  magnetic energy density is much larger than that of the mass density,
being as much as $\sim 10^7.$ The  plots are typical of
what is found   once the turbulence has reached a  quasi steady state.

\begin{figure}
\centerline{
\epsfig{file=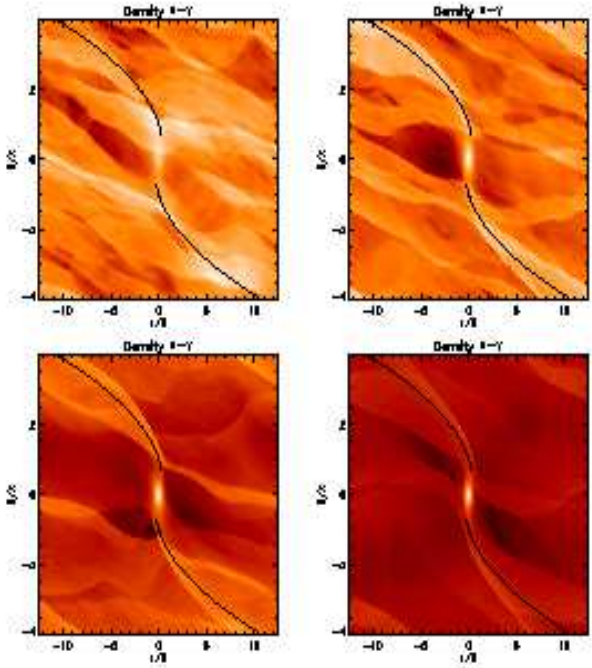,height=14.cm,width=10.cm,angle =0} }
\caption[]
{
Plots corresponding to those
in figure \ref{fig3} but for models Bb1, Bb2, Bb3, Bb4.
A similar sequence is indicated.
}
\label{fig11}
\end{figure}

\noindent
\begin{figure}
\centerline{
\epsfig{file= 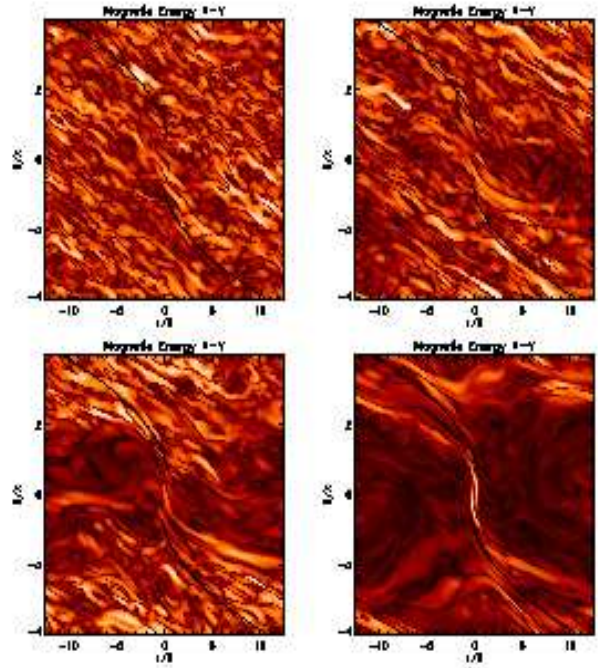,height=14.0cm,width=10.cm,angle =0} }
\caption[]
{Plots corresponding to those
in figure \ref{fig3a} but for models Bb1, Bb2, Bb3, Bb4.
A similar sequence is indicated.
}
\label{fig12}
\end{figure}

When a protoplanet is placed in the centre of the box waves are excited
that cause  outward angular momentum transport. These waves also
have an associated Reynolds' stress (e.g. Papaloizou \& Lin 1984)  and 
may affect the underlying turbulence.

The  volume averaged  total  stress parameter
$\langle \alpha \rangle$ is plotted
as a function of time for models Ba1, Ba2, Ba3, and Ba4 in
figure \ref{fig10}. The plots commence  from the moment when the protoplanet was introduced.
The introduction causes the excitation of a wave and an increase in the Reynolds'
stress which in turn increases the total stress.
This is particularly noticeable in simulations Ba3 and Ba4
which have  $M_p R^3/(M_* H^3) = 1.0$ and $2.0 $ respectively 
and so  have large enough protoplanets to form a gap.
 This behaviour was also found in the simulation of a $5$ Jupiter mass planet
in a disc with $H/R =0.1$  presented in paper II and in the global simulations
described in section~\ref{global-res}.
After a sharp initial rise the volume averaged  Reynolds' stress here  tends to decrease
as a gap is opened and material is moved away from the protoplanet.
The  much less noisy  volume averaged 
magnetic stresses   are also plotted in figure \ref{fig10}.
These  behave similarly  for each of the simulations
Ba1, Ba2, Ba3 and Ba4 but they then  separate at about time $500.$
Although simulation Ba4 which has the highest protoplanet mass
also tends to have the highest volume averaged magnetic stress, there is no other
such correlation between magnetic stress and protoplanet mass and so this may not be significant.

\begin{figure}
\centerline{
\epsfig{file= 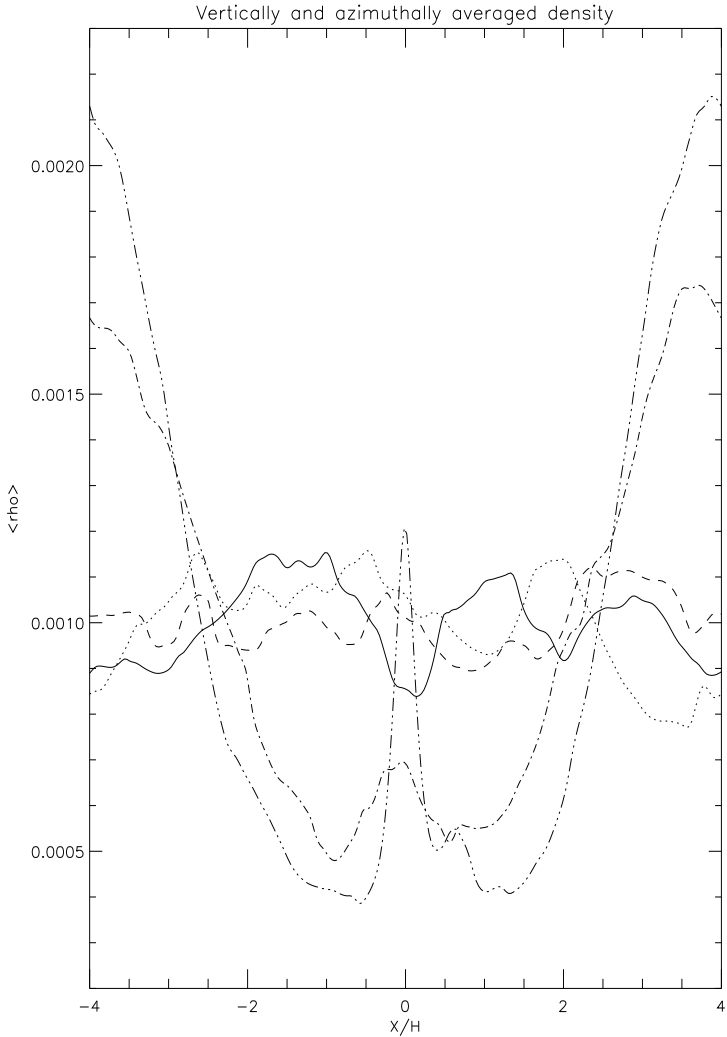,height=14.cm,width=10.cm,angle =0} }
\caption[]
{Density averaged over $y$ and $z$ plotted against $x$ for  simulations
Ba0 full curve, Ba1  dotted curve,
Ba2 dashed curve, Ba3 dot dashed curve, and Ba4 triple dot dashed curve,
taken near the end of the simulations.
These had $M_p R^3/(M_* H^3) = 0.1, 0.3, 1.0$ and $2.0 $ respectively.
The plots show that $M_p R^3/(M_* H^3)$ increases,
material is accreted by the protoplanet and is pushed toward the radial
boundaries as a gap is formed. Such a gap is noticeable in simulations
Ba3 and Ba4 but not in simulations Ba1 and Ba2 which have a very similar
behaviour of the mean density to that found in simulation Ba0 which had no
protoplanet. Although there are turbulent fluctuations
the same structure persists once a quasi steady state is achieved.}
\label{fig6}
\end{figure}

\begin{figure}
\centerline{
\epsfig{file= 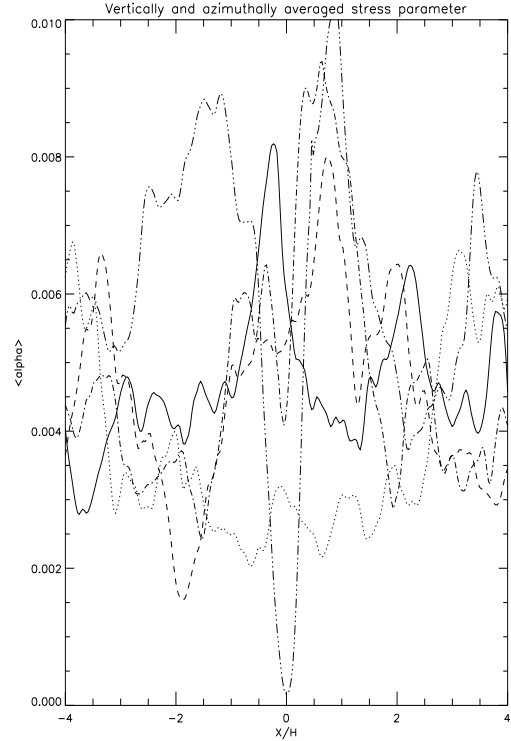,height=10.cm,width=7.cm,angle =0}}
\caption[]
{ The stress parameter averaged over $y$ and $z$ plotted against $x$ for  simulations
Ba0 full curve, Ba1  dotted  curve,
Ba2 dashed curve, Ba3 dot dashed curve, and Ba4 triple dot dashed curve,
taken near the end of the simulations. Although there are erratic fluctuations,
the variation in $\langle \alpha \rangle$  among the different simulations is typically
less than that   found in the density in the gap region. This is consistent
with a general reduction of magnetic energy density  and stress in that region
when a gap is formed.
}
\label{fig7}
\end{figure}

Although, as we have indicated above, strong fluctuations are indeed the rule,
there are  definite  behavioural trends as the protoplanet mass increases.
To illustrate these, density contour plots in a typical 
$(x,y$) plane for simulations
Ba1, Ba2 Ba3 and Ba4 
near the end of the simulations are shown in figure \ref{fig3}.
It can be seen that as $M_p R^3/(M_* H^3)$ increases, the wake becomes more prominent.
It can be  seen  that this wake approximately coincides with that
given by equation (\ref{crcv}) derived from the discussion about characteristic rays.
The wake is the first observable manifestation of the presence of the 
protoplanet
and can be just traced in simulation Ba1 for which $M_p R^3/(M_* H^3) = 0.1.$
This is well before any effects can be seen in the magnetic energy density contours
or a gap forms.
In figure \ref{fig3a}  we plot the magnetic energy density  contours
corresponding to the density plots in figure \ref{fig3}.

As $M_p R^3/(M_* H^3)$ increases and a gap forms, the  magnetic energy  generally
decreases in the gap region. Where it remains it tends to be concentrated in regions
near the high density wakes. This tendency was also apparent in the global simulation of
paper II and is also observed in the global simulations G4 and G5 described
in section~\ref{global-res}. 

To examine the effect of lengthening the box in the  azimuthal or
$y$ direction, we  carried out simulations
Bb1, Bb2, Bb3  and Bb4 which had the same resolution as Ba0, Ba1, Ba2, Ba3  and Ba4
but were extended by a factor of $2$ to $8\pi$ in $y.$
However, because of the obvious increased computational demands, these simulations 
were run for shorter times. Each was begun by taking simulation Ba0 at a time $216,$
extending it to $8\pi$ in $y$ by using the appropriate
periodicity condition and inserting protoplanets
with masses given in table \ref{table1}.
Mass density and magnetic energy density plots are plotted in
 figures \ref{fig11}
and \ref{fig12} respectively. A similar behavioural trend to that found for the
shorter boxes is noted.

To illustrate the gap formation process the 
density averaged over $y$ and $z$ is  plotted against $x$ for  simulations
Ba0, Ba1,
Ba2, Ba3  and Ba4  at a moment
near the end of the simulations in figure \ref{fig6}.
As $M_p R^3/(M_* H^3)$ increases,
material is pushed toward the radial
boundaries by the action of the wave stresses
and a gap is formed. 
In addition it is apparent that there is some accretion
onto the protoplanet in the sense that some material settles
onto it becoming bound. The horizontally and vertically
averaged density profiles associated with Ba0, 
 Ba1 and Ba2  indicate no sign of a gap with the variations between them
similar  to what would be seen for any one of them at different times.
However a persistent gap is clearly visible for runs Ba3 and Ba4 for
which $M_p R^3/(M_* H^3) > 1.$

The  horizontally and vertically averaged 
stress parameter corresponding to the plots in figure \ref{fig6}
is plotted against $x$ for simulations
Ba0, Ba1,
Ba2, Ba3 and Ba4 in figure \ref{fig7}.
Although there are erratic fluctuations,
the range of variation in $\langle \alpha \rangle$    between the simulations is typically
less than that  found in the density when there is a gap. This is consistent
with a tendency for there to be 
a general reduction of magnetic energy and stress in a gap region
that roughly scales with the decrease in gas density and pressure there.

We have also compared the structure of the gap near the end
of the simulation Ba4 with that obtained from a laminar
simulation with no magnetic field but  uniform anomalous Navier Stokes
viscosity corresponding to the time average value
of $\alpha =0.008$  obtained from run Ba0
which had no planet. We emphasize that such a time average
has no clear physical connection with the behaviour of simulation
Ba4 nonetheless we perform the comparison for general interest.

\begin{figure}
\centerline{
\epsfig{file=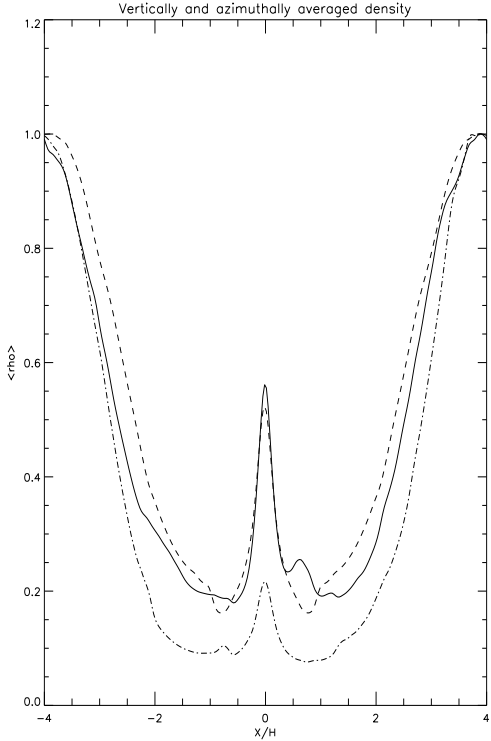,height=10.cm,width=7.cm,angle =0}}
\caption[]
{ The density averaged over $y$ and $z$ plotted against $x$ for  simulation
Ba4 full curve, the laminar disc simulation
with $\alpha =0.008,$ dashed  curve and the inviscid  laminar  simulation,
dot dashed curve,
taken near the end of the simulations.
The inviscid case has the deepest gap and the case
with $\alpha =0.008$ the narrowest.
}
\label{visccomp}
\end{figure}

In figure \ref{visccomp} we plot the azimuthally and vertically
averaged density profile at the end of simulation Ba4 together
with plots obtained from laminar disc runs with
$\alpha =0.008$ and no viscosity which were run for the same times
with the perturbing planet which had $M_p R^3/(M_* H^3) = 2.$
These have been scaled to have the same magnitude at the box edges. 

As expected, the inviscid non magnetic run produces a deeper and
wider gap than the others. The run with the imposed
$\alpha =0.008$ produces a somewhat narrower gap than
the MHD run Ba4 which in this way behaves
as if it had lower viscosity. This  effect was also found for
the global simulation with 5 Jupiter mases  in paper II.
However, in our case there is slightly more material
accreted onto the planet than the laminar viscous case.

\begin{figure}
\centerline{
\epsfig{file= 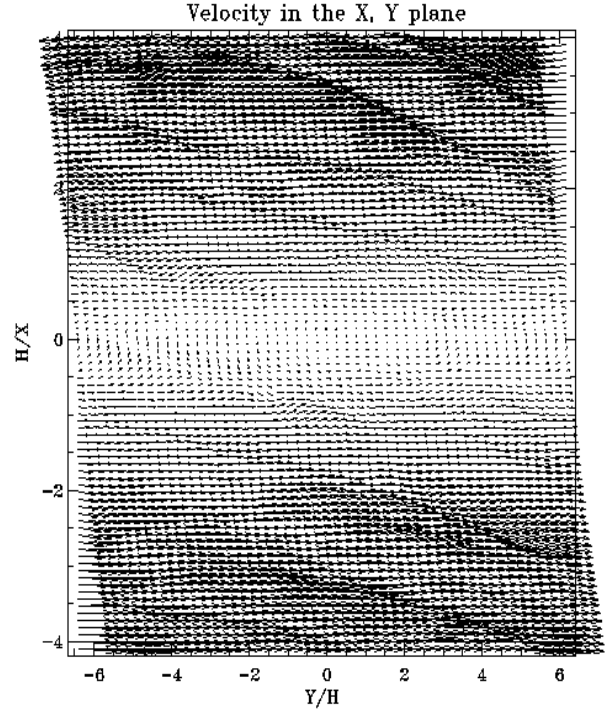,height=15.cm,width=10.cm,angle =0} }
\caption[]{ Velocity vectors in a typical $(x,y)$ plane   near the end of simulation
Ba1 with $M_p R^3/(M_* H^3) = 0.1.$ The perturbation due to the protoplanet is
too weak for the existence of horse shoe trajectories that do not cross
the whole $y$ domain  to be manifest.
}
\label{fig4}
\end{figure}

\begin{figure}
\centerline{
\epsfig{file=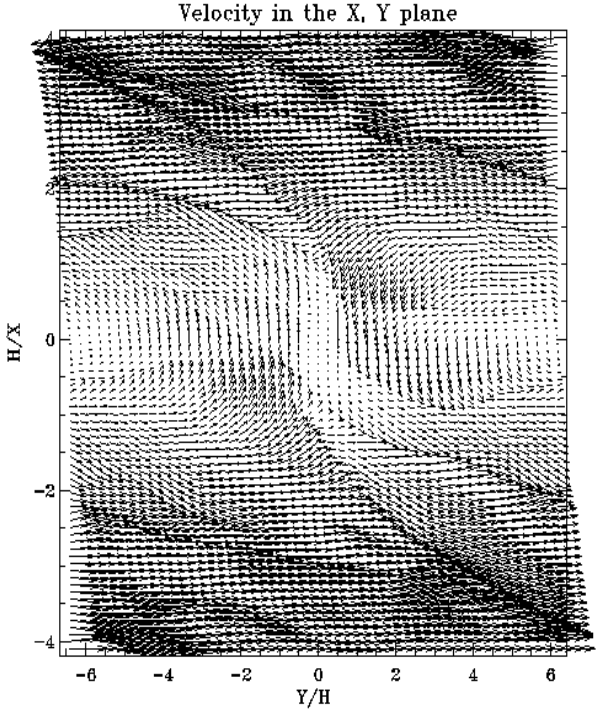,height=15.cm,width=10.cm,angle =0} }
\caption[]
{ Velocity vectors in a typical $(x,y)$ plane for simulation
Ba3 with $M_p R^3/(M_* H^3) = 2.$ The perturbation due to the protoplanet is
 sufficient for the existence of horse shoe trajectories
restricted to a subset of the $y$ domain to be manifest.}
\label{fig5}
\end{figure}

 Finally we examine the flow in the coorbital region
to check for the appearance of circulating horse shoe like trajectories in that region.
Velocity vectors in  typical $(x,y)$ planes for simulations Ba1, weakly perturbed
with $M_p R^3/(M_* H^3) = 0.1$  and 
Ba3, strongly perturbed with  $M_p R^3/(M_* H^3) = 2$ are presented
in figures \ref{fig4}
and \ref{fig5} respectively.  Circulating trajectories
are apparent in the case of Ba3 indicating the existence of a coorbital dynamics.
However, the flow pattern appears complex with circulating eddies and multiple shocks.
Furthermore we find no indication of a circulating flow around the protoplanet
within the Hill sphere.
This may be because of a magnetic braking effect already discussed in paper II
(also see discussion in section~\ref{global-res}).

\subsection{Global Simulations} \label{global-res}
In this section we describe the results obtained from the global
simulations. We begin by discussing the underlying turbulent disc model,
before describing the interaction between protoplanets and global 
turbulent discs.

\subsubsection{Global Turbulent Disc Model}
The physical parameters of the global disc model are described in 
section~\ref{global-setup}. The initial disc model had a $\pi/4$ azimuthal
domain, with an imposed zero-net flux toroidal field that
varied sinusoidally with radius. The initial value of
the ratio
of the total magnetic energy to the volume integrated pressure,
$ \int {\bf B}^2/(8\pi)dV/ \int PdV  = \langle 1/\beta \rangle=0.032$
The time evolution of this quantity is shown in figure~\ref{fig14}.
It is clear that the initially high magnetic energy decreases as the
initially imposed magnetic field undergoes reconnection,
and $\langle 1/\beta \rangle$ reaches a saturated value of 
$\langle 1/\beta \rangle \simeq 0.011$ once the turbulence reaches a statistical
steady state. 

\begin{figure}
\centerline{
\epsfig{file= 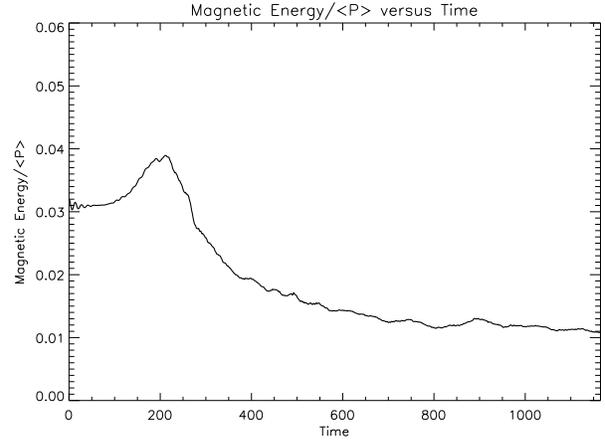,width=\columnwidth} }
\caption[]
{This figure shows the evolution of $\langle 1/\beta \rangle$ for the global
disc model. }
\label{fig14}
\end{figure}

Figure~\ref{fig15} shows the time evolution of the volume averaged 
Shakura-Sunyaev stress
parameter $\alpha$ defined by equation~\ref{alpha}. The solid line represents
the Maxwell stress, the dotted line represents the 
Reynolds stress, and the dashed line represents the sum of these.
It is clear that as 
the disc becomes unstable to the MRI, the stresses increase initially,
but then decrease to a saturated state corresponding to 
$\alpha \simeq 7 \times 10^{-3}$. 

\begin{figure}
\centerline{
\epsfig{file=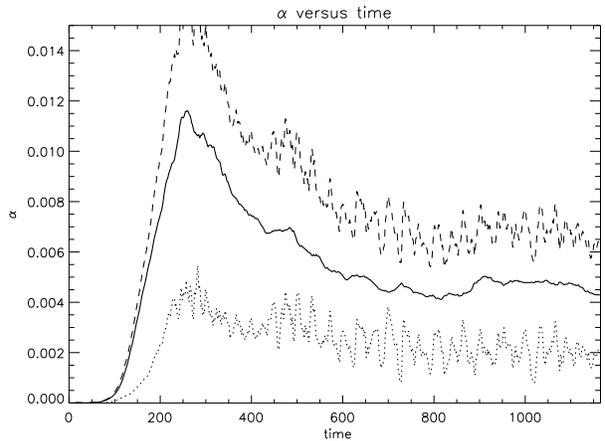,width=\columnwidth} }
\caption[]
{This figure shows the evolution of $\langle \alpha \rangle$ for the global
disc model. The solid line represents
the Maxwell stress, the dotted line represents the
Reynolds stress, and the dashed line represents the total stress.}
\label{fig15}
\end{figure}

The radial distribution of $\alpha$, 
time averaged for a short period of time (one orbit at $r=1$) is shown in 
figure~\ref{fig16}. As expected this is a rapidly varying quantity as a function
of both space and time due to the turbulence (see also paper I).

\begin{figure}
\centerline{
\epsfig{file=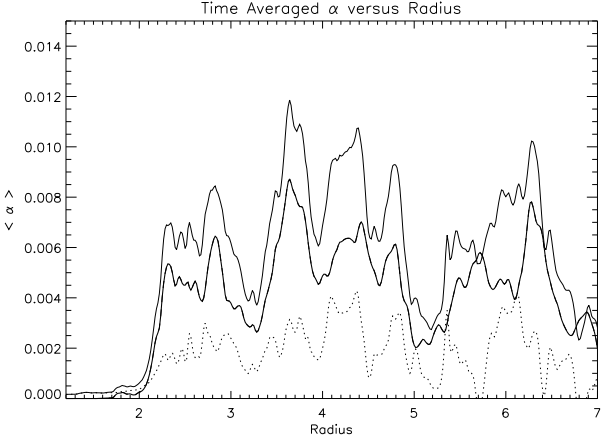,width=\columnwidth} }
\caption[]
{This figure shows the radial profile of $\langle \alpha \rangle$ for the 
global
disc model. The heavy solid line represents
the Maxwell stress, the dotted line represents the
Reynolds stress, and the thin solid line represents the total stress.}
\label{fig16}
\end{figure}

The final state of the disc model shown in figures~\ref{fig14} -- \ref{fig16}
was used as the initial condition for simulations that examine the 
interaction between turbulent discs and embedded protoplanets.
Models G1 --  G4 used full $2 \pi$ azimuthal domains that were constructed
by patching eight copies of the $\pi/4$ turbulent disc model together.
Model G5, which had a restricted azimuthal domain of $\pi/2$, 
consisted of two copies of the relaxed, turbulent disc model patched together,
with periodic boundary conditions in the azimuthal direction being employed.
These simulations are described below.

\subsubsection{Global Models with Embedded Protoplanets}
In this section we present the results from simulations labeled
as G1 -- G5 in table~\ref{table2}. The aim is to examine how the
disc properties change as a function of planet mass (or alternatively
as a function of $(M_p/M_*)/(H/r)^3$), moving from a situation
where the protoplanet induces linear perturbations in the disc (e.g. run G1)
through to a scenario in which the planet perturbation is nonlinear 
(e.g. run G5). In particular, we are interested in how the appearance of
the spiral density waves vary, how the morphology of the magnetic field
changes in the vicinity of the planet, when gap formation arises
in a global disc model, and the impact of the protoplanet on the
velocity field in the disc. We are also interested in how the results
of similar local and global simulations compare.

When discussing the results of the global
simulations we will concentrate on the
runs G1, G2, G3, and G5. Run G4 adopted the exact same physical parameters
as run G5, except that the azimuthal domain in G5 ran from
$0$ to $\pi/2$, rather than being a full $2 \pi$ domain. This means that
we were able to run simulation G5 for $\simeq 60$ planetary orbits because
of the lower computational cost, whereas run G4 was only run for $\simeq 11$
planetary orbits. We therefore give the results of run G5 higher
prominence in our discussion, whilst noting that the results obtained
in run G4 are in good qualitative agreement with G5.

Figure~\ref{fig17} shows contour plots of the midplane density distribution
in the vicinity of the protoplanet for the runs G1, G2, G3, and G5, 
respectively. In the first panel it is clear that the presence of the planet
in the disc is essentially undetectable since the density perturbations
induced by the turbulence are of higher amplitude than the spiral waves
excited by the protoplanet. The value of $(M_p/M_*)/(H/r)^3$ in this case
is 0.03, with the protoplanet mass corresponding to $\sim 3$ Earth masses
(assuming a solar mass central star).
The second panel, corresponding to run G2, shows that the presence 
of the protoplanet is just about detectable, but the amplitude of 
perturbations due to the turbulence still exceed the 
spiral waves induced by the protoplanet. The protoplanet mass in this case
is equivalent to $\sim 10$ Earth masses. The third panel corresponds to run 
G3, for which the protoplanet mass is equivalent to $\sim 30$ Earth masses.
The presence of the protoplanet in this case is clearly detectable, with
the spiral waves being of similar amplitude to the turbulent wakes.
The fourth panel shows the results from run G5, in which the protoplanet mass
corresponds to 3 Jupiter masses. In this case the disc--protoplanet
interaction leads to gap formation since the interaction is strongly
nonlinear ($(M_p/M_*)/(H/r)^3) \simeq 8$ in this case). Overall
the results here are in very good agreement with those obtained in the
shearing box runs, in that they show the same consistent
pattern of behaviour with increasing values of $(M_p/M_*)/(H/r)^3$.

We next consider the effect of the protoplanet on the magnetic field.
Figure~\ref{fig18} shows contour plots of $|{\bf B}|^2$
for the runs G1, G2, G3, and G5. Strong variation in magnetic field
strength arises due to the turbulence, as found in the shearing box runs,
and the plots show that the magnetic
field is largely unaffected by the presence of the embedded protoplanets in
runs G1, G2, G3. However, the last panel, corresponding 
to run G5, clearly shows that
the magnetic field strength is increased near the spiral shock waves induced
by the more massive planet, in agreement with results obtained in paper II
and those obtained by the shearing box runs. In general, the magnetic energy 
and stress decrease in the gap region as the density and pressure decrease, 
but the
disordered, turbulent magnetic field close to the 
planet becomes ordered and compressed in the spiral wakes, leading
to a significant increase in the magnetic stress there.
This effect is shown more clearly in figure~\ref{fig19} which shows a snapshot
of the magnetic field vectors in run G5. This illustrates how the field
becomes compressed and ordered as the gas travels through the spiral shocks,
and also shows how the magnetic field is advected into the protoplanet
Hill sphere with gas that accretes onto the protoplanet.
The effect that the field modification has on the vertically and azimuthally
averaged magnetic stress is illustrated
by figure~\ref{fig20} which shows the time averaged contribution
to $\alpha$ from the magnetic stress as a function of radius.
It is clear that the compression of
the field along with reduction of the density and pressure in the gap leads to
an enhanced $\alpha$ value there.

\begin{figure*}
\centerline{
\epsfig{file=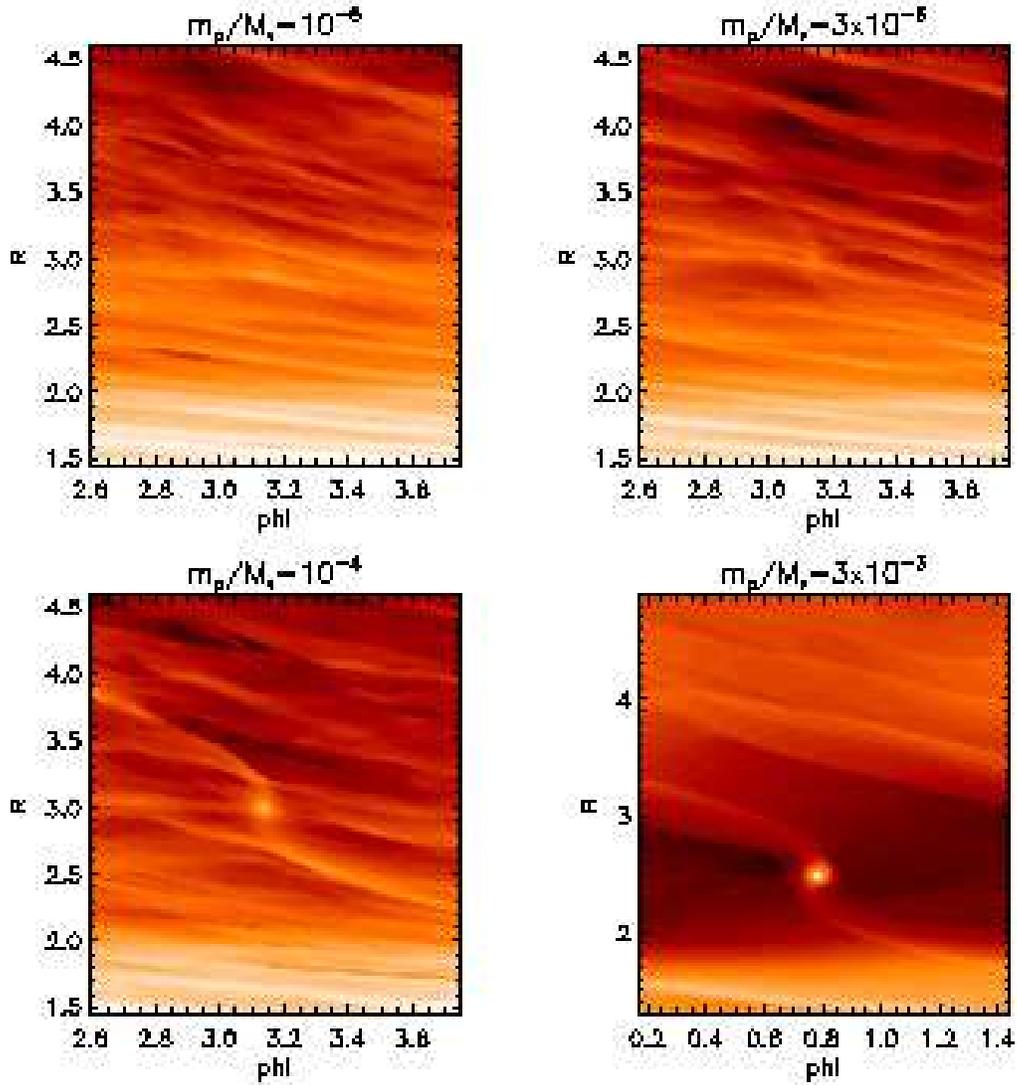,width=\textwidth} }
\caption[]
{This figure shows the midplane density distribution of runs G1, G2, G3, G5,
respectively. The progression of increasing planet mass leads to 
a clear trend in the disturbance experienced by the disc, with the spiral waves 
becoming increasingly apparent. Run G5 leads to the formation of a clear gap.}
\label{fig17}
\end{figure*}

\begin{figure*}
\centerline{
\epsfig{file=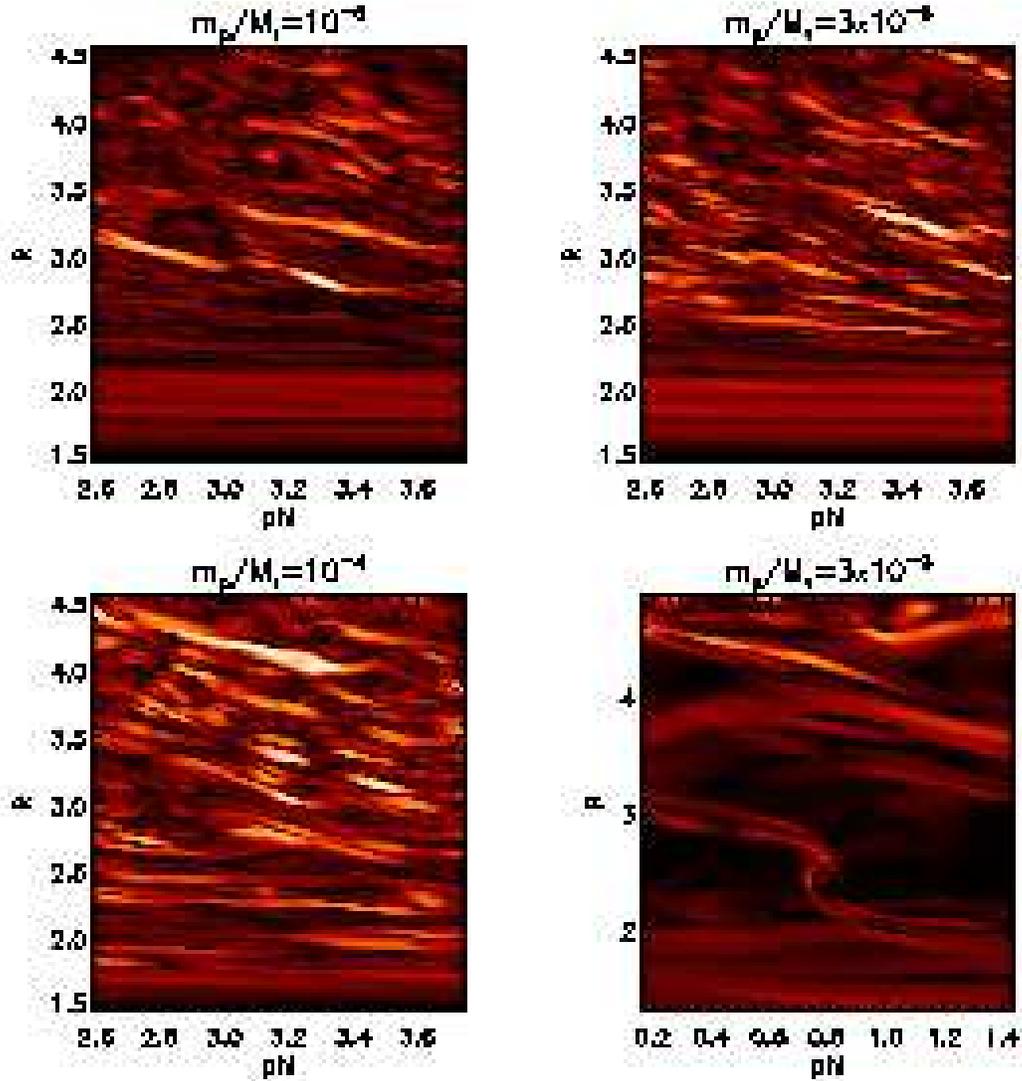,width=\textwidth} }
\caption[]
{This figure shows the distribution of $|B|^2$ in the disc midplane
for the runs G1, G2, G3, G5,
respectively. For runs G1, G2, G3 the protoplanet make no discernible
perturbation to the magnetic field structure. For run G5 he magnetic field
strength is increased in the wakes generated by the planet.}
\label{fig18}
\end{figure*}

\begin{figure}
\centerline{
\epsfig{file=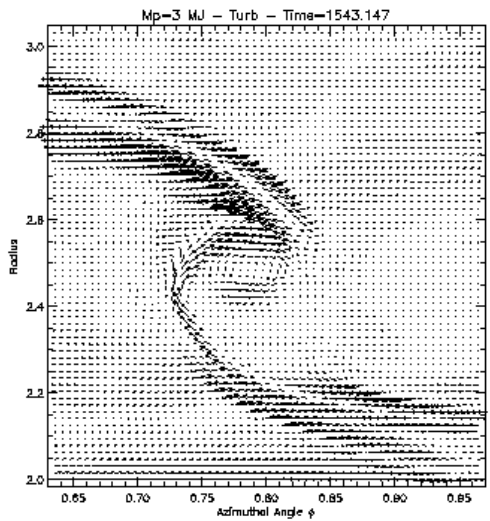,width=\columnwidth} }
\caption[]
{This figure shows magnetic field vectors in the vicinity of the protoplanet
for run G5. The compression and ordering of the field due to the
spiral shocks is apparent, as is the advection of magnetic flux 
into the planet Hill
sphere.}
\label{fig19}
\end{figure}

\begin{figure}
\centerline{
\epsfig{file=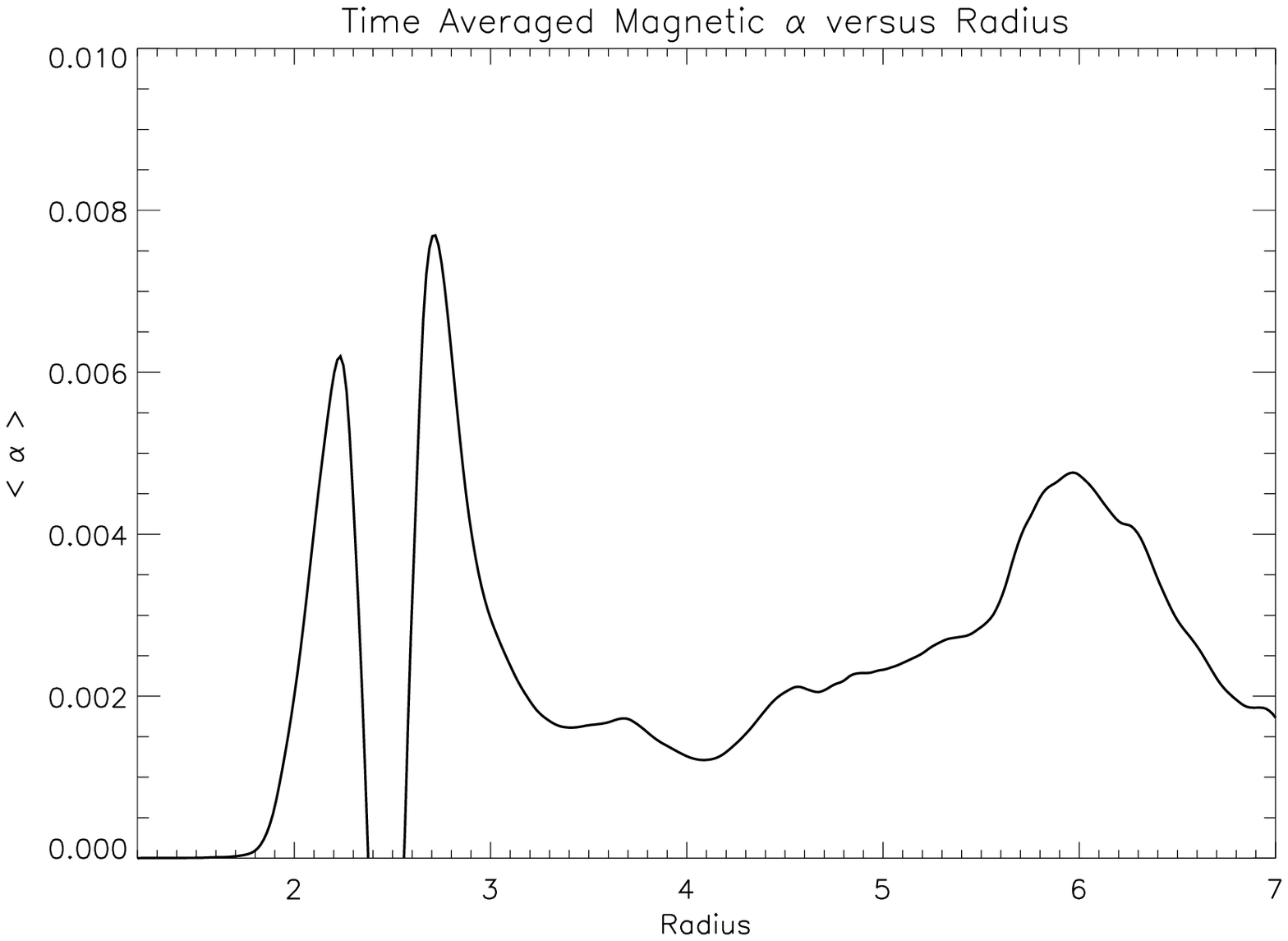,width=\columnwidth} }
\caption[]
{This figure shows a time average of the magnetic contribution to $\alpha$
for run G5. The time averaging occurred for 18 planet orbits. The increase
in $\alpha$ in the gap region occurs because of the field
concentration in the spiral wakes described in the text.}
\label{fig20}
\end{figure}

We now consider how the presence of the protoplanet affects the radial density
distribution in the disc. Figure~\ref{fig21} shows the azimuthally averaged
midplane density distribution for the runs G1, G2, G3, and G5, as well as the
initial $1/r$ density profile.
This plot shows that some accretion though the disc has occurred during its
initial relaxation to a final turbulent state. The lower mass protoplanets
in runs G1, G2, and G3 have essentially no effect on the underlying
density structure, but a deep gap is formed in run G5.
This is in line with the expectations discussed
in section~\ref{Scaling} since $(M_p/M_*)/(H/r)^3 < 1$ for G1, G2, and G3, but
$(M_p/M_*)/(H/r)^3 \simeq 8$ for run G5, and is in very good agreement with the
shearing box runs.

\begin{figure}
\centerline{
\epsfig{file=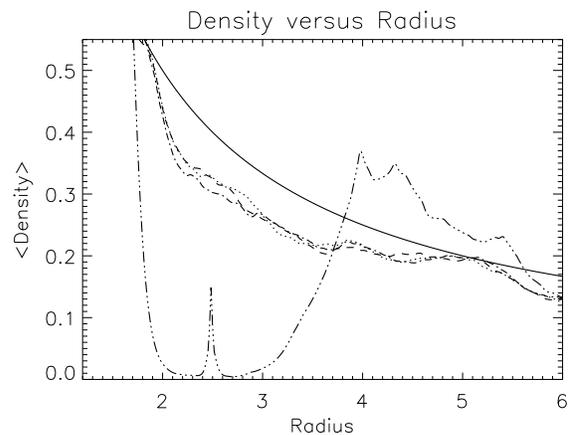,width=\columnwidth} }
\caption[]
{This figure shows the radial density distribution for the
runs G1, G2, G3, and G5. The solid line shows the initial
$1/r$ density profile of the unperturbed disc. The dotted line
corresponds to run G1, the dashed line to G2, and the dash-dotted line to G3.
It is clear that there is no gap forming in these cases.
The dash-dot-dot-dotted line corresponds to run G5 which shows clear
gap formation.}
\label{fig21}
\end{figure}

We now turn to the question of how the global energetics of the turbulence
and the stresses arising from it are affected by the presence of the 
protoplanets. In the case of the lower mass planets represented by
runs G1, G2, and G3 the global magnetic energy of the simulations
in units of the volume integrated pressure is essentially
unaffected. This is in agreement
with the results obtained for a massive 5 Jupiter mass planet embedded 
in a  disc
with $H/r=0.1$ presented in paper II and the shearing box runs.
Figure~\ref{fig22} illustrates this 
with a plot of $\langle 1/\beta \rangle$ versus time for runs G1, G2, and G3.
It is clear that the magnetic energy remains close
to its original starting value, with the level of variation no
larger than the fluctuations normally associated with the turbulent 
energy. Simulation G3 was initiated with an underlying
disc model that had relaxed for a slightly shorter period
of time than models G1 or G2, which were initiated with identical underlying 
turbulent disc models. Although the qualitative evolution of the global
magnetic energy is the same in the three cases, it is also
interesting to compare the detailed evolution of $\langle 1/\beta \rangle$ for
models G1 and G2. Figure~\ref{fig22} shows that initially the
evolution is almost identical. 
However, after $\sim 100$ time units they clearly begin
to diverge due to the `chaotic' nature of the turbulence, as
noted by Winters, Balbus, \& Hawley (2003b). A similar result is
obtained in the shearing box runs. Nonetheless
the global properties of the disc models remain similar, even though the
local details change. This also has an impact on the way in which the local
statistics of the turbulence are affected by the protoplanet, which in turn
affects the detailed evolution of the orbit of 
the planet. These issues are 
discussed in greater length in the accompanying paper IV.

In the same way that the magnetic energy is largely unaffected by the presence
of the low mass protoplanets, the turbulent stresses are also unaffected.
This is illustrated by figure~\ref{fig23} which shows the various
contributions to $\alpha$ as a function of time for run G2 and
should be compared with figure~\ref{fig15}. Similar
plots are obtained for runs G1 and G3.

\begin{figure}
\centerline{
\epsfig{file=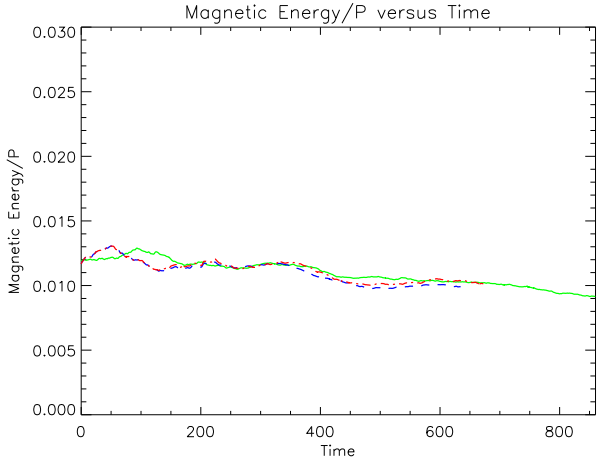,width=\columnwidth} }
\caption[]
{This figure shows the evolution of the magnetic energy in units of
the volume integrated pressure for runs G1, G2, and G3. 
It is clear that the presence of the planets has little effect on 
the turbulent dynamo as the magnetic energy remains close to its initial 
value. Variations occur with amplitudes no greater 
than those associated with the usual turbulent fluctuations.}
\label{fig22}
\end{figure}

\begin{figure}
\centerline{
\epsfig{file=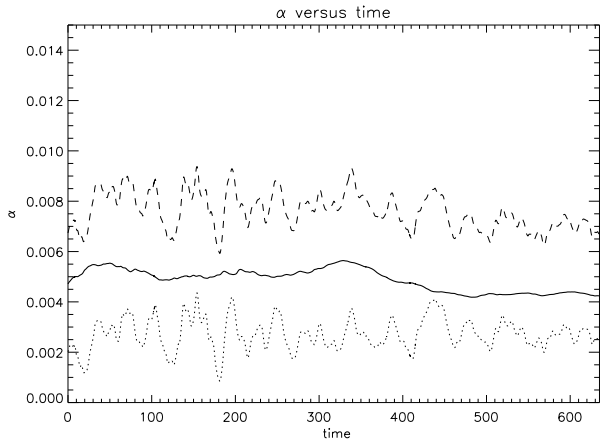,width=\columnwidth} }
\caption[]
{This figure shows the evolution of the stress parameter $\alpha$ for the run 
G2. The upper line represents the total stress, the middle line the Maxwell 
stress, and the lowest line the Reynolds stress. The total 
$\alpha \simeq 7 \times 10^{-3}$.}
\label{fig23}
\end{figure}

The picture is a little different when we consider run G5. The time evolution
of $\langle 1/\beta \rangle$ is shown in figure~\ref{fig24}, which shows that
the evolution of the magnetic energy appears to decrease during the simulation,
before reaching a minimum value of $\langle 1/\beta \rangle \simeq 0.005$
beyond which it decreases no further.
Such a decrease did not occur during the other global runs (G1-G3), or
in the calculation of a massive planet in a turbulent disc in paper II.
At the present time it is unclear whether the turbulent
energy is being affected by the protoplanet in such a manner that the turbulent
dynamo is operating less efficiently due to the ongoing planetary perturbation,
or whether the decrease is the result
of a large but temporary fluctuation
induced by inserting the planet instantaneously in
the turbulent disc. The former scenario is conceivable
when one considers that the azimuthal domain in run G5 is $\pi/2$ so that
the fluid experiences the (strong) perturbation of the planet four times
more frequently than it would do in a full $2 \pi$ disc. Such a
strong and frequent perturbation may be able to affect the underlying
dynamo in such a way as to reduce the magnetic energy. The shearing box runs
in general do not support this picture, but run Bb4 did also show
a reduction in magnetic energy once the planet was inserted.
It is interesting to note that the results of Winters, Balbus, \& Hawley (2003a)
also contain a suggestion that the saturation level of the MHD turbulence
may be affected by the presence of a giant planet.

\begin{figure}
\centerline{
\epsfig{file=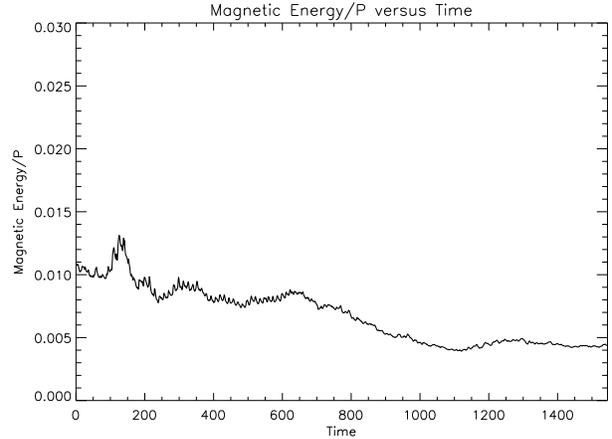,width=\columnwidth} }
\caption[]
{This plot shows the evolution of the magnetic energy, in units of the
volume integrated pressure, for run G5. This quantity decreases initially
from $\sim 0.01$ to $\sim 0.045$ before leveling off.} 
\label{fig24}
\end{figure}

We now consider the impact of the embedded protoplanets on the
velocity field of the disc, and specifically the point
at which the horseshoe orbits are clearly established.
Figure~\ref{fig25} shows the fluid trajectories for two runs.
The upper panel corresponds to a global calculation in which a 
protoplanet with $M_P/M_*=10^{-4}$ is embedded in a laminar disc (i.e.
the protoplanet is the same as that in run G3).
Close inspection of this plot shows that the horseshoe 
trajectories are established
in this case. The lower panel shows the velocity field
for run G3, where it is apparent that the horseshoe trajectories
are disrupted by the turbulent velocity field. In other words the turbulence
determines the fluid trajectories in this case, and the gravitational
potential
due to the planet is unable to establish the horseshoe orbits in the coorbital
zone that are
obtained in a laminar disc model. 

\begin{figure}
\centerline{
\epsfig{file=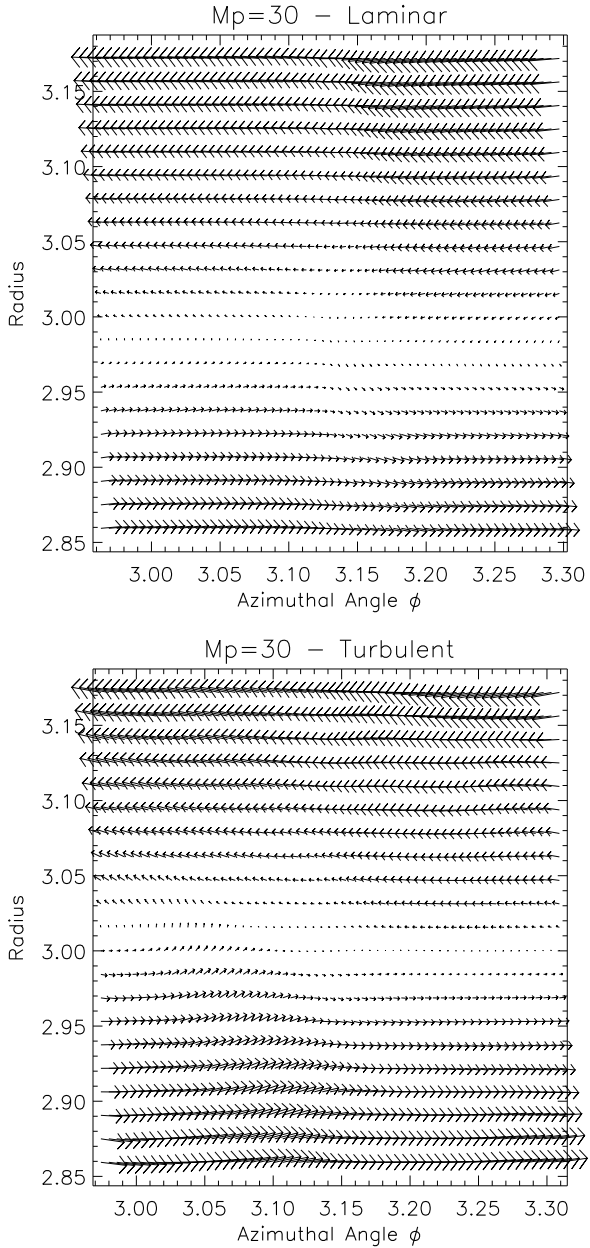,width=\columnwidth} }
\caption[]
{This plot shows the fluid trajectories for a laminar run with
$M_P/M_*=10^{-4}$ (upper panel) and for run G3 (lower panel).
The usual horseshoe orbits can be seen in the upper panel,
but the turbulence appears to dominate the trajectories in the lower panel.
Note that the protoplanet is located at $(r_p,\phi_p)=(\sim 3,\pi)$.
We note that plotting the velocity field for run G3 at different times
leads to significant variation in the appearance of the velocity field
because of the turbulence, but that the flow morphology in the laminar case
is essentially independent of time once established.}
\label{fig25}
\end{figure}

Figure~\ref{fig26} shows the fluid trajectories for run G5.
The horseshoes orbits in this case are very clearly defined, in
agreement with the shearing box run Ba4. In this model the
circulating region around the planet in the Hill sphere is also clearly visible.
The effect of the turbulence on the velocity field in
such a strongly perturbed model is essentially indiscernible
in the near vicinity of the planet. 

In paper II the circulating region around the protoplanet in the Hill
sphere was found to be 
disrupted, and this was tentatively ascribed to magnetic braking
caused by magnetic linkage between the circumplanetary disc
and the protostellar disc (see also figure~\ref{fig19}). A similar
situation was also found in the shearing box runs Ba3 and Ba4,
where the usual circulating region in the Hill sphere was found to be
absent. In these runs the gravitational softening adopted was  $b \simeq 0.3H$,
which is quite large, and results in gas that accretes into the Hill sphere
forming an `atmosphere' that is largely pressure supported but with some
angular momentum. In this case the removal of angular momentum
will lead to a reduction of the spin of the atmosphere, as observed.
The softening used in run G5 was $b = 0.1H$,
and so the formation of a rotationally supported 
circumplanetary disc is more pronounced. The 
removal of angular momentum in this case
should allow further gas accretion into 
the Hill sphere without a modification of the rotation profile.
An MHD simulation should therefore result in more material
accreting into the planetary Hill sphere than occurs in a non magnetised
run, if magnetic braking is indeed important. In order to test
this we performed a simulation that was similar to run G5, except that
magnetic fields were neglected and an $\alpha$ viscosity was employed with
$\alpha=7 \times 10^{-3}$ in a laminar disc model.
This laminar disc calculation was evolved for
an identical amount of time to run G5 ($t=1543.14$ $\Omega^{-1}$, equivalent
to $\simeq 62.12$ planetary orbits), and resulted in less mass
being accreted into the planetary Hill sphere than occurred for
the magnetised run, as shown in figure~\ref{fig28}.
This suggests that magnetic fields in turbulent discs will play a signifcant
role in determining the gas accretion rate onto forming gas giant planets.

\begin{figure}
\centerline{
\epsfig{file=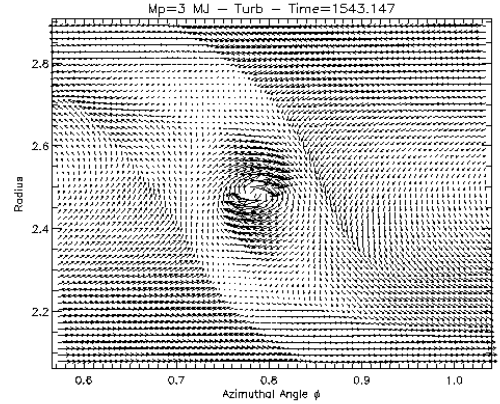,width=\columnwidth} }
\caption[]
{This plot shows the fluid trajectories in the vicinity of the
protoplanet for run G5.
The horseshoe orbits are clearly visible, as is the circulating region
within the Hill sphere.}
\label{fig26}
\end{figure}

\section {Discussion} \label{S5}

In this paper we have performed both global cylindrical disc simulations
and local shearing box simulations 
of protoplanets interacting with
a disc undergoing MHD turbulence with zero net flux fields.
The aim has been to extend the results obtained in a previous
paper (paper II) to a wider range of protoplanet masses
and conditions of perturbation of the surrounding disc.

\begin{figure}
\centerline{
\epsfig{file=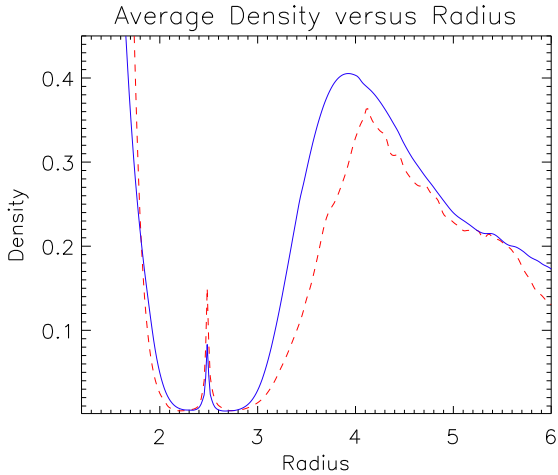,width=\columnwidth} }
\caption[]
{This plot shows the azimuthally averaged midplane density
as a function of radius for run G5 (dashed line), and for a similar run with a
viscous but laminar disc in which $\alpha=7 \times 10^{-3}$ (solid line).
The plots correspond to a run time of $t=1543.14$. It is clear that
more gas has accumulated within the Hill sphere of run G5 than for
the laminar disc run, suggesting that magnetic braking of the circumplanetary
disc is playing a significant role in modulating the accretion rate onto
the planet.}
\label{fig28}
\end{figure}

Global simulations are naturally
more realistic but their computational demands mean that only very few can be carried out. 
The advantage of local shearing box calculations is that
more runs can be done at higher resolution even though larger boxes
than are normally considered (e.g. Brandenburg et al. 1995; Hawley et al. 1995)
are required in order to adequately
contain the response to a perturbing protoplanet.

Another advantage is that for zero net flux fields
there exists  a natural scaling indicating that results
depend only on the parameters  $M_p' = M_p R^3/(M_* H^3)$
and $Y/H.$ 
Using simple dimensional considerations of the condition
for non linear response and balance of viscous and tidal torques
we estimated a simple condition for gap formation
given by equation~\ref{condres} in the form
$M_p R^3/(M_* H^3) > \rm{ max}  (C_t, 0.07(Y/H)).$
For the conditions of the local and global simulations $(Y = \pi R)$
we have performed both here and in paper II,  this always 
results in a condition that $M_p R^3/(M_* H^3)$ exceed a number
of order unity (i.e. the thermal condition mentioned
by Lin \& Papaloizou 1993).

The pattern of behaviour we have found from our simulations
is the same for both local and global simulations (which show
excellent agreement in the results obtained for the local
interaction between protoplanet and turbulent disc) and it 
agrees with the notion described above.
As $M_p R^3/(M_* H^3)$ is increased to $\sim 0.1$
the presence of the protoplanet is first manifest with the appearance
of the well known trailing wake that can 
be identified with a characteristic ray emanating from the protoplanet.
This wake appears to be well defined even though its form may vary
erratically due to the turbulent fluctuations. At this stage the magnetic
field in the disc is relatively undisturbed by the planet, and the usual
horseshoe orbits in the coorbital region are poorly defined
because the fluid trajectories are still dominated by the turbulence.

When $M_p R^3/(M_* H^3)$ exceeds  a number around
unity a gap starts to develop inside
which the magnetic energy density is on average less than that in the rest
of the  fluid. However, in the gap region it becomes 
concentrated in the high density wakes
that are generated by the protoplanet. At this point the horseshoe
orbits in the coorbital zone become apparent as the gravity of the protoplanet
dominates the turbulence.
A region in which fluid circulates about the protoplanet
may be formed in some cases and this may contain a  magnetic field
that circulates about the protoplanet and links to the
wakes, leading to the possibility of magnetic braking
of the circumplanetary disc.
In this situation the two sides of the disc tend to separate as the gap forms,
signaling the onset of type II migration (e.g. Ward 1997).

The nature of the migration induced by disc interactions
is of key importance during the later stages of planet formation.
This has yet to be considered for a disc with MHD turbulence for low mass
embedded protoplanets. The simulation presented in paper II
indicates that migration at the expected rate occurs when a gap is fully formed
in a turbulent disc, in agreement with type II migration theory.
We comment here that 
the migration rate induced by the interaction is a complex issue
on account of the strong turbulent fluctuations, especially for embedded 
protoplanets.
This issue is examined in a companion paper (paper IV).

\subsection{Acknowledgements} 
The computations reported here were performed using the UK Astrophysical
Fluids Facility (UKAFF).


{}
\end{document}